\documentclass{nature}
\usepackage{amsmath, amssymb, times, mathrsfs, hyperref, array, bbm, ulem,}
\usepackage{graphicx}
\usepackage{xcolor}

\hypersetup{colorlinks=false}

\title{Induced superconducting correlations in the quantum anomalous Hall insulator}
\author{Anjana Uday$^{1,5}$, Gertjan Lippertz$^{1,2,5}$, Kristof Moors$^{3}$, Henry F. Legg$^{4}$, Andrea Bliesener$^{1}$, Lino M. C. Pereira$^{2}$, A. A. Taskin$^{1}$, Yoichi Ando$^{1}$}

\makeatletter
\let\saved@includegraphics\includegraphics
\AtBeginDocument{\let\includegraphics\saved@includegraphics}
\renewenvironment*{figure}{\@float{figure}}{\end@float}
\makeatother

\begin{document}

\maketitle

\begin{affiliations}
 \item Physics Institute II, University of Cologne, Z\"ulpicher Str. 77, 50937 K\"oln, Germany
 \item KU Leuven, Quantum Solid State Physics, Celestijnenlaan 200 D, 3001 Leuven, Belgium
 \item Peter Gr\"unberg Institute 9, Forschungszentrum J\"ulich \& JARA J\"ulich-Aachen Research Alliance, 52425 J\"ulich, Germany
 \item Department of Physics, University of Basel, Klingelbergstrasse 82, CH-4056 Basel, Switzerland
 \item These authors contributed equally
\end{affiliations}


\begin{abstract}
Inducing Cooper pairing in a thin ferromagnetic topological insulator in the quantum anomalous Hall state (called quantum anomalous Hall insulator, QAHI) is a promising way to realize topological superconductivity with associated chiral Majorana  edge states. 
However, finding evidence of superconducting proximity effect in a QAHI has proven to be a considerable challenge due to inherent experimental difficulties. Here we report the observation of crossed Andreev reflection (CAR) across a narrow superconducting Nb electrode contacting the chiral edge state of a QAHI, evinced by a negative nonlocal voltage measured downstream from the grounded Nb electrode. This is an unambiguous signature of induced superconducting pair correlation in the chiral edge state. Our theoretical analysis demonstrates that CAR processes of the chiral edge are not strongly dependent on the nature of the superconductivity that mediates them. Nevertheless, the characteristic length of the CAR process is found to be much longer than the superconducting correlation length in Nb, which suggests that the CAR is in fact mediated by superconductivity induced on the QAHI surface. The approach and results presented here provide a foundation for future studies of topological superconductivity and Majorana physics, as well as for the search for non-Abelian zero modes.
\end{abstract}

\begin{figure}
\centering
\includegraphics[width=\textwidth, trim={0.5cm 0.5cm 0.5cm 0cm}, clip]{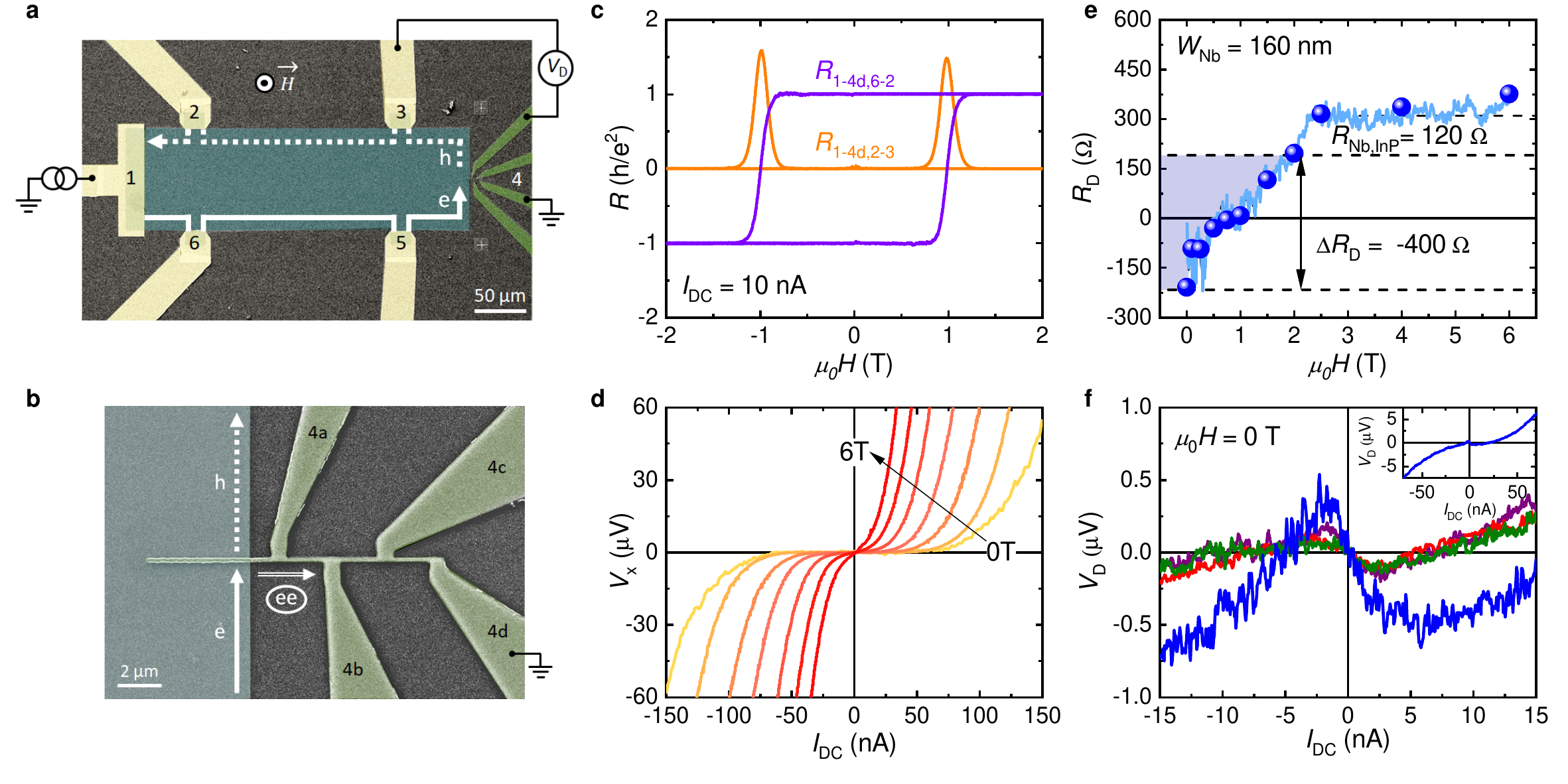}
\caption{\linespread{1.05}\selectfont{}
{\bf Crossed Andreev reflection (CAR) across the quantum anomalous Hall edge state. } 
\textbf{a,} False-colour scanning electron microscopy (SEM) image of device A including the measurement schematics. The superconducting Nb electrode (green) and the Ti/Au normal electrodes (yellow) contact the V-doped (Bi$_{x}$Sb$_{1-x}$)$_2$Te$_3$ thin film (cyan). For an upward, out-of-plane magnetization ($M>0$), the chiral 1D edge state propagates counter-clockwise along the sample edge; the voltage $V_{\rm D}$ between contact 3 and 4 gives the downstream resistance $R_\text{D} \equiv V_{\rm D}/I_{\rm DC}$. \textbf{b,} Magnified image of the 160-nm-wide Nb electrode shown in panel \textbf{a}. The white arrows schematically show the CAR process. \textbf{c,} Magnetic-field dependence of the 4-terminal resistances, showing the QAHE with vanishing longitudinal resistance $R_\text{1-4d,2-3} = 0$ and quantized transverse resistances $R_\text{1-4d,6-2} = h/e^2$ at 25 mK. \textbf{d,} Current-voltage ($I$-$V$) characteristics of the 4-terminal longitudinal voltage $V_x$ at 17 mK in various applied magnetic fields $H$ from 0 to 6 T in 1-T step. The breakdown current decreases with increasing $H$. \textbf{e,} Light blue line shows the downstream resistance $R_\text{D}$ continuously measured as a function of $H$ from 0 to 6 T with $I_\text{DC}=2$ nA at 25 mK. Blue symbols represent the slopes of the $I$-$V$ characteristics at $I_{\rm DC}$ = 0 at discrete magnetic fields (see Supplementary Note 3), which give confidence in the negative $R_\text{D}$ indicative of CAR. As the superconductivity in Nb is suppressed with increasing $H$, $R_\text{D}$ increases by 520 $\Omega$, which consists of the normal-state Nb resistance (120 $\Omega$, marked by a dashed line) and the CAR contribution $\Delta R_\text{D} \simeq -400$ $\Omega$ (marked by blue shade). \textbf{f,} Multiple $I$-$V$ curves for the downstream voltage $V_\text{D}$ measured in 0 T at 17 mK for different magnetic-field-sweep histories. The magnitude of the negative slope at $I_\text{DC}$ = 0 depends on the history, see Supplementary note 4 for details. Inset shows an $I$-$V$ curve up to $\pm70$ nA dominated by the current-induced breakdown of the QAHE.}
\label{fig:QAHE+R_D}
\end{figure}

Inducing superconducting (SC) correlations using the SC proximity effect in the one-dimensional (1D) edge state of a two-dimensional (2D) topological system would lead to exotic topological superconductivity hosting non-Abelian anyons \cite{Fu2009, Qi2010, Lindner2012, Meng2012, Clarke2013, Vaezi2013, Vaezi2014} and hence has been experimentally pursued in a couple of systems. In the case of the 1D helical edge state of a 2D topological insulator (TI), the induced SC correlations have been detected in Josephson junctions \cite{Hart2014, Wiedenmann2016}. The SC correlations in the quantum Hall edge states are less trivial due to the chiral nature of the edge and large magnetic fields required, but strong evidence has been obtained in terms of the crossed Andreev reflection (CAR) \cite{Clarke2014, Hou2016, Beconcini2018, Galambos2022} or the formation of Andreev edge states \cite{Hoppe2000, Chtchelkatchev2007, Khaymovich2010} causing a negative nonlocal potential in the downstream edge \cite{Lee2017, Zhao2020, Hatefipour2022, Gul2022}. In the CAR process, an electron in the chiral edge entering a grounded SC electrode creates a Cooper pair by taking another electron from the other side of the electrode, causing a hole to exit into the downstream edge (see Fig.\ref{fig:QAHE+R_D}b). This hole is responsible for the negative nonlocal voltage observed experimentally \cite{Lee2017,Gul2022}. Importantly, SC correlations are induced in the chiral edge state through the CAR process. Very recently, the CAR process was observed even in the fractional quantum Hall edge states \cite{Gul2022}, which are an interesting platform to create parafermions obeying rich non-Abelian statistics \cite{Mong2014, Alicea2016}. 

In this context, the SC proximity effect in the quantum anomalous Hall insulator (QAHI), which is a ferromagnetic TI showing the quantum anomalous Hall effect (QAHE), is highly interesting. If the 1D edge state of a QAHI can be proximitized, one could create a non-Abelian zero mode by coupling two counter-propagating edges via the CAR process through a superconductor~\cite{Vaezi2013, Clarke2014, Lee2017}.
If, on the other hand, the 2D surface of the QAHI is proximitized, chiral Majorana edge state may show up \cite{Qi2010, Wang2015}. The edge vortex in the chiral Majorana edge state is a non-Abelian zero mode \cite{Beenakker2019a} and can be used for building flying topological qubits that transfer information between stationary qubits \cite{Beenakker2019b, Adagideli2020, Bauerle2018}. 
Hence, proximitized QAHI is an interesting platform for Majorana physics and non-Abelian zero modes. However, no clear evidence has been reported for the SC proximity effect in a QAHI~\cite{Kayyalha2020, Shen2020, Thorp2022}. 

The QAHI can be realized by doping Cr or V to a very thin (typically $\lesssim$ 10 nm thickness) film of the three-dimensional (3D) TI material (Bi$_x$Sb$_{1-x}$)$_2$Te$_3$ in which the chemical potential is fine-tuned into a magnetic gap opened at the Dirac point of the surface states as a result of a ferromagnetic order \cite{Yu2010, Chang2013, Chang2015}. Hence, a QAHI is insulating not only in the 3D bulk but also in the 2D surface. Inducing SC correlations in bulk-insulating TIs is much more difficult than in bulk-conducting TIs \cite{Ghatak2018}, and this is one of the reasons for the lack of clear evidence for the SC proximity effect in a QAHI. In fact, a recent work reported the observation of Andreev reflection (AR) in a metallic regime of a magnetic TI film, but when the sample was in the QAHI regime, there was no evidence for any Andreev process \cite{Kayyalha2020}. Another work in this context \cite{Shen2020} had a device structure not optimal for detecting the relevant Andreev process. Recent experiments on the quantum Hall system found a robust signature of CAR even at the spin-polarized $\nu=1$ filling factor\cite{Lee2017, Gul2022}, which appears similar to the QAHI edge; however, an important difference is that the QAHI edge is not fully spin polarised\cite{Zhang2016}. In the present work, we have successfully observed the signature of CAR with a narrow Nb finger-electrode (down to $\sim$160 nm width) contacting the QAHI edge. The finger-width dependence of the CAR signal gives the characteristic length of the CAR process that is much longer than the SC coherence length of Nb, which suggests that it is not the superconductivity in the Nb electrode but is the proximity-induced pairing in the QAHI beneath the Nb that is mediating the CAR process.

\begin{flushleft}
{\bf Nonlocal detection of the crossed Andreev reflection}
\end{flushleft}
\vspace{-8mm}

Our samples are Hall-bar devices of V-doped (Bi$_x$Sb$_{1-x}$)$_2$Te$_3$ \cite{Lippertz2022} contacted by superconducting Nb-electrodes with widths ranging from 160 to 520 nm. Figures \ref{fig:QAHE+R_D}a-b show false-colour scanning electron microscopy (SEM) images of device A which had the narrowest Nb electrode (contact 4). All other contacts are made of Ti/Au with contact resistances of a few Ohms (see Supplementary Note 1). The 1D chiral edge state propagates in the counter-clockwise direction for an upward, out-of-plane magnetization ($M>0$). For the configuration shown in Fig.\ref{fig:QAHE+R_D}a, we set a DC current to flow between contacts 1 and 4d; namely, a voltage is applied to the normal-metal contact 1 and the SC contact is grounded. 

In Ref.~\citenum{Kayyalha2020}, the AR of the electrons in the 2D ``bulk'' states of a magnetic TI film in the metallic regime was observed in devices similar to Fig. \ref{fig:QAHE+R_D}a, but here we probe the SC correlations in the 1D chiral edge state of the QAHI. For our purpose, confirmation of the dissipationless edge transport without the contribution of the 2D bulk is essential. In fact, the longitudinal resistance $R_\text{xx}$ ($=R_\text{1-4d,2-3}$ measured between contacts 2--3 with the current between 1--4d) vanishes in our devices, while the transverse resistance $R_\text{yx}$ ($=R_\text{1-4d,6-2}$ measured between contacts 6--2) is quantized to $h/e^2$ without the need of electrostatic gating, as shown is Fig. \ref{fig:QAHE+R_D}c. Note that a breakdown of the zero-resistance state occurs when the current exceeds a critical current \cite{Kawamura2017, Fox2018, Rodenbach2021 ,Fijalkowski2021, Lippertz2022, Qiu2022, Zhou2022}, and the zero-resistance region is observed to shrink with increasing magnetic fields \cite{Bottcher2019,Bottcher2020}, as shown in Fig. \ref{fig:QAHE+R_D}d. This fragility of the QAHI state against current makes it difficult to estimate the contact transparency using current biasing \cite{Lee2017, Kayyalha2020}.



The CAR process converts an incoming electron with energy $eV$ smaller than the SC gap $\Delta$ to a hole carrying a potential of $-V$ in the downstream edge (Fig. \ref{fig:QAHE+R_D}b), which is detected at contact 3 as the downstream voltage $V_{\rm D}$ with respect to the grounded SC contact 4a. Here, downstream refers to the chiral direction of the edge state, see Figs. \ref{fig:QAHE+R_D}a-b. 
In addition, there is a finite probability that an upstream electron tunnels directly to the downstream as an electron carrying a positive potential $V$; this is called co-tunneling (CT) and it competes with the CAR process in the nonlocal transport \cite{Galambos2022, Beconcini2018, Hou2016}.
The downstream resistance $R_\text{D} \equiv V_{\rm D} / I_{\rm DC}$ observed in this configuration consists of
\begin{equation}
R_\text{D} = R_\text{QAHI} + R_\text{Nb,InP} + R_\text{SC-QAHI},
\label{eq:R_D}
\end{equation}
where the resistance of the QAH film $R_\text{QAHI}$ is zero for low probe currents below the breakdown, $R_\text{Nb,InP}$ is the resistance of the Nb section lying on the InP wafer between the film edge and the SC contact 4a (which is zero when the Nb is superconducting), and $R_\text{SC-QAHI}$ is the SC-QAHI interface resistance including the nonlocal CAR/CT contribution. Note that the present setup is a 3-terminal configuration and the contact resistance of the Nb electrode always gives a finite contribution to $V_{\rm D}$. An external magnetic field is not required for the realization of the QAHE, enabling us to examine $R_\text{SC-QAHI}$ as a function of the applied magnetic field from 0 T up to the upper critical field $H_{\rm c2}$ of superconducting Nb.
This is an important difference from the previous studies of the SC proximity effect in the quantum Hall edge states \cite{Lee2017, Zhao2020, Hatefipour2022, Gul2022}. 
Note, however, that even in zero applied magnetic field, the magnetization of the QAHI induces vortices in the Nb electrode, which create subgap states and allow incident electrons to be dissipated without the Andreev mechanism \cite{Lee2017,Tang2022,Schiller2023}. 
Due to the chiral nature of the edge state, no AR occurs into the upstream edge.

Figure \ref{fig:QAHE+R_D}e shows the magnetic-field dependence of $R_\text{D}$ for device A with a Nb electrode width $W_\text{Nb}$ of 160 nm, measured with $I_\text{DC}=2$ nA (see Supplementary Notes 3--4 for additional data). Below $\sim$1 T, the downstream resistance is negative, signalling the CAR process across the Nb electrode. This is the most significant result of this work and demonstrates that superconducting correlations are induced in the chiral edge state across the SC finger via CAR processes in our devices. As the magnetic field is increased, $R_\text{D}$ gradually turns positive and saturates as the superconductivity is lost in the Nb electrode. 
The change in the nonlocal downstream resistance due to the suppression of the CAR/CT process is calculated as $\Delta R_\text{D} \equiv -[R_\text{D}(H > H_\text{c2})-R_\text{D}(H < H_\text{c2})-R_\text{Nb,InP}]$. When $\Delta R_\text{D}$ is negative (positive), the CAR (CT) process is dominant. We estimate $\Delta R_\text{D} \approx -400$~$\Omega$ after subtracting the contribution of the normal-state Nb resistance $R_\text{Nb,InP} \approx 120$ $\Omega$ (see Fig. \ref{fig:QAHE+R_D}e and Supplementary Note 2). 
Thus, the magnetic-field dependence of $R_{\rm D}$ allows us to elucidate the positive contribution of the contact resistance and conclude that the CAR process is contributing $\Delta R_{\rm D} \simeq -400$ $\Omega$, which is much larger than the measured negative $R_{\rm D}$. This $\Delta R_{\rm D}$ corresponds to about 3\% of the maximum negative downstream resistance $-h/2e^2$ expected for prefect crossed Andreev reflection (see Supplementary Note 8).


To give confidence that the negative $R_{\rm D}$ is not just a result of voltage fluctuations, the $I$-$V$ characteristics for the downstream voltage $V_\text{D}$ in 0 T are shown in Fig. \ref{fig:QAHE+R_D}f. The slope in the zero-current limit (which also gives $R_\text{D}$) is reproducibly negative for all the measured curves for different magnetic histories, even though the magnitude of $R_\text{D}$ changes with the magnetic history (see Supplementary Note 5), which is probably caused by a change in the disorder profile. Notice that a small nonreciprocal contribution is superimposed on the $I$-$V$ curves in Fig. \ref{fig:QAHE+R_D}f, which is likely a result of the nonreciprocity of the 1D chiral edge transport in the QAHI \cite{Yasuda2020, Baumgartner2022}. Moreover, while a negative slope is observed for $|I_\text{DC}| \lesssim 3$ nA, the breakdown of the QAHE (causing $R_\text{QAHI} > 0$) dominates the downstream voltage at high current values (see Fig. \ref{fig:QAHE+R_D}f inset).

\begin{flushleft}
{\bf Temperature and finger-width dependences of the downstream resistance}
\end{flushleft}
\vspace{-8mm}

\begin{figure}
\centering
\includegraphics[width=0.8\textwidth]{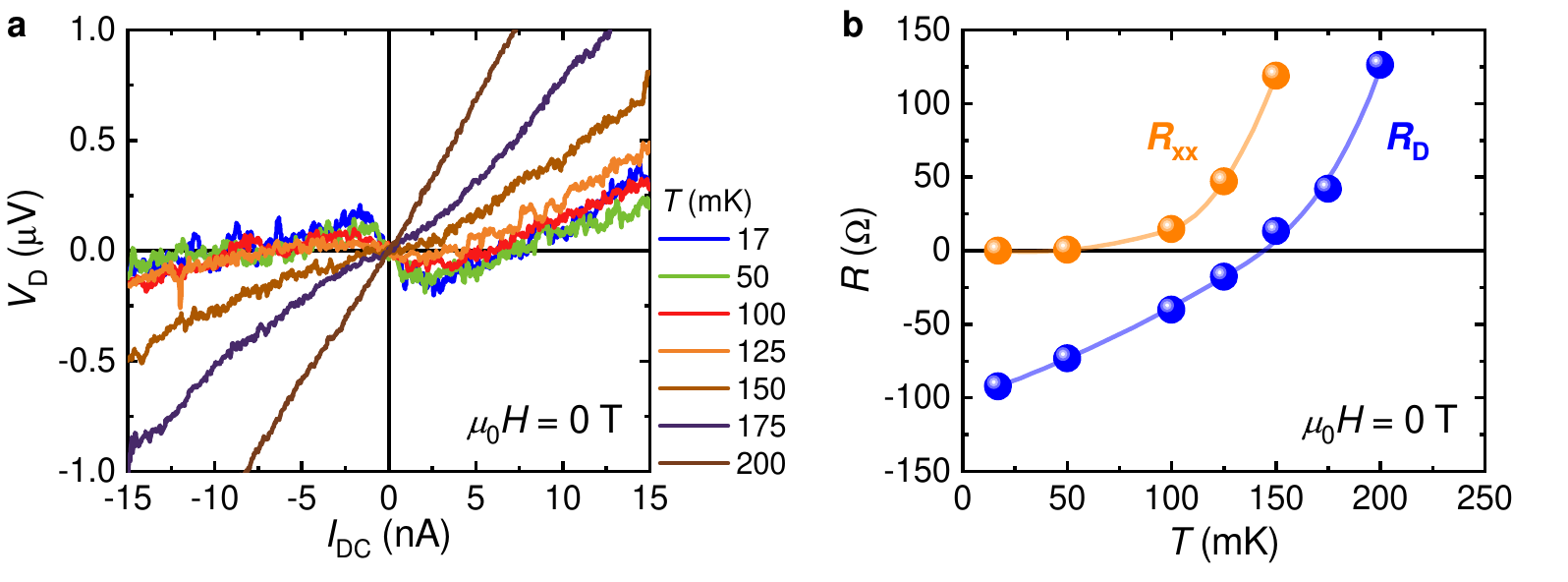}
\caption{\linespread{1.05}\selectfont{}
{\bf Temperature dependence of the downstream potential in device A. } 
\textbf{a,} Plots of $V_{\rm D}$ vs $I_{\rm DC}$ at different temperatures measured with the setup shown in Fig.\ref{fig:QAHE+R_D}a. \textbf{b,} Temperature dependencies of $R_\text{D}$, extracted from the $I$-$V$ curves in panel \textbf{a} at $I_{\rm DC}$ = 0 (blue), and the 4-terminal longitudinal resistance $R_\text{xx}$ (orange). Above 50 mK, $R_\text{xx}$ deviates from zero, indicating that the dissipationless transport of the QAHE is lost. Consequently, the 2D bulk resistance eventually dominants $R_\text{D}$ at $T \gtrsim$ 100 mK.}
\label{fig:IV_T}
\end{figure}

Figure \ref{fig:IV_T}a shows the plots of $V_{\rm D}$ vs $I_{\rm DC}$ taken at different temperatures up to 200 mK, and the $R_\text{D}$ extracted in the zero-current limit is plotted in Fig. \ref{fig:IV_T}b as a function of temperature. $R_\text{D}$ is negative up to $\sim$125 mK, which is much smaller than the critical temperature $T_\text{c}$ of Nb. This is understood by looking at the temperature dependence of the 4-terminal longitudinal resistance $R_{xx}$, which starts to deviate from zero above $\sim$50 mK (see Fig. \ref{fig:IV_T}b), a typical behaviour of the QAHI samples available today \cite{Kawamura2017, Fox2018, Rodenbach2021, Fijalkowski2021, Lippertz2022}. Hence, the CAR contribution in $R_\text{D}$ is masked by the contribution of $R_\text{QAHI}$ at $T >$ 50 mK.

To further analyze the nonlocal CAR process across the Nb contact, we investigated $R_{\rm D}$ for devices with different Nb-finger widths up to 520 nm. The magnetic-field dependence of $R_{\rm D}$ in device B with $W_\text{Nb}$ = 235 nm is shown in Fig. \ref{fig:Width_Dep}a (see Supplementary Note 7 for the data of devices C-E having larger widths). The estimated Nb finger resistance $R_\text{Nb,InP}$ is also shown for comparison. Notice that the increase in $R_\text{D}$ coincides with the suppression of the superconductivity in Nb. The $R_{\rm D}$ value fluctuates around zero in this  $W_\text{Nb}$ = 235 nm sample when Nb is superconducting, which is an indication that the CAR contribution happened to be nearly of the same magnitude as the normal contact resistance, so that the resulting $R_\text{SC-QAHI}$ is around zero. Note that the simple AR can only account for a factor of two reduction in the interface resistance \cite{Blonder1982, Kayyalha2020} and cannot explain $R_\text{D}$ going to zero. We estimate $\Delta R_\text{D}  = -70$ $\Omega$ for sample B (see Fig. \ref{fig:Width_Dep}a).

For comparison, we show in Fig. \ref{fig:Width_Dep}c the data for a $W_\text{Nb}$ = 160 nm sample (device F) which was fabricated several months after the film was grown. The aging of the film caused a large normal contact resistance and the CAR contribution $\Delta R_\text{D}$ cannot make $R_\text{SC-QAHI}$ to become negative or zero, even though the width of this sample is the same as that of sample A. Using the estimated $R_\text{Nb,InP} \simeq$ 100 $\Omega$, we obtain the normal-state $R_\text{SC-QAHI}$ of about 420 $\Omega$ and $\Delta R_\text{D} \simeq -170$ $\Omega$ for this sample F. We note that devices A--E are of higher quality because they were fabricated on a fresh QAHI film immediately after the MBE growth.

As shown in Figure \ref{fig:Width_Dep}b, although no negative downstream resistance was observed for devices B--E with wider Nb electrodes than device A, a finite negative $\Delta R_\text{D}$ was always obtained up to $W_\text{Nb} \simeq$ 500 nm. For device A ($W_\text{Nb}$ = 160 nm), as already mentioned, different values of the negative $R_\text{D}(H < H_\text{c2})$ were obtained for different magnetic-field sweeps due to changing disorder profiles (see Supplementary Note 5) and they are included in Fig. \ref{fig:Width_Dep}b as individual data points. 
One can see in the inset of Fig. \ref{fig:Width_Dep}b that, on average, the magnitude of $\Delta R_\text{D}$ is exponentially suppressed with increasing $W_\text{Nb}$. A fit to $\Delta R_\text{D}=R_0\exp(-W_\text{Nb}/\xi_\text{CAR})$ gives $R_0 \approx -750$ $\Omega$ and the characteristic length of the CAR process of $\xi_\text{CAR} \approx 100$ nm. This is much longer than the SC coherence length of dirty Nb, i.e. $\sqrt{\xi_\text{BCS} l_\text{mfp}} \approx 30$ nm (with $\xi_\text{BCS} = \hbar \nu_F^S / \pi \Delta$, $\nu_F^S = 1.37 \times 10^{6}$ m/s, $\Delta = 1.2$ meV, and $l_\text{mfp} \approx 3$ nm)\cite{Mermin1976,Zaysteva2020}.

\begin{flushleft}
{\bf Origin of crossed Andreev reflection}
\end{flushleft}
\vspace{-8mm}

At first sight, the chiral edge state of a QAHI appears similar to the $\nu=1$ state of the quantum Hall insulator\cite{Lee2017,Gul2022}. However, the spin-polarised nature of the $\nu=1$ state necessitates, for instance, a superconductor with strong spin-orbit coupling or a nonuniform magnetic field distribution in order for CAR processes to occur\cite{Lee2017,Gul2022,Galambos2022}. In contrast, the edge state of the QAHI considered in our work is a superposition of spin states on the top and bottom surfaces. Furthermore, when brought into proximity with a superconductor, the resultant doping of the TI surface\cite{Legg2022,Russmann2022} will lead to an induced superconductivity that inherently has strong spin-orbit coupling.  In a simple approximation and at zero-momentum, the wavefunction of the chiral edge in the $x$-direction of a QAHI takes the form~\cite{Yu2010,Shen2020} (see Supplementary Note 9)
\begin{equation}
\Psi(y)=f(y) \left(\left| t \uparrow \right\rangle+\chi \left|b \downarrow \right\rangle \right),
\end{equation}
where $f(y)\sim e^{-y/\lambda}$ is an exponentially decaying envelope function with length-scale $\lambda$ determined by the Fermi-velocity and magnetic gap opened in the 2D topological surface states \cite{Shen2020}, $ \left|t\uparrow \right\rangle$ ($\left|b \downarrow \right\rangle$) denotes the spin-up (spin-down) state in the top (bottom) surface of the QAHI, and $\chi$ specifies the top/bottom anisotropy. In the isotropic case, $\chi=1$, the edge state has no net spin-polarisation and the CAR process will not be hindered as long as the SC finger is narrow enough.  Superconducting pairing across the finger will be only slightly suppressed by a partial spin-polarisation of the edge states, which may arise in realistic situations \cite{Shen2020}. 


We now turn to possible scenarios by which SC correlations could be introduced into the edge states via CAR processes, starting first with a scenario where the SC finger defines a trivial SC region, such that no chiral Majorana (CM) edge state can form. The Nb finger is itself trivial and, under certain circumstances, the induced proximitised SC state in the TI surface can also be trivial \cite{Wang2015}. Apart from the factors discussed above, this scenario is essentially identical to that of the $\nu=1$ quantum Hall state that has previously been extensively discussed\cite{Lee2017,Galambos2022, Beconcini2018, Hou2016, Schiller2023, Kurilovich2022}. We note, however, that the disordered nature of the QAHI surface would cause the Andreev edge state to become diffusive and results in an equal mixture of electrons and holes, such that the Andreev edge state will not contribute to $R_{\rm D}$. Taking $\xi_s$ to be the SC coherence length of the mediating superconductor, the CAR processes induce superconducting correlations across the SC finger and gives rise to negative $V_{\rm D}$ when the finger width $W_{\rm SC}$ is shorter than $\xi_s$. When $W_{\rm SC} \gg \xi_s$, the transport along the Andreev edge state dominates over CAR.

\begin{figure}
	\centering
	\includegraphics[width=\textwidth]{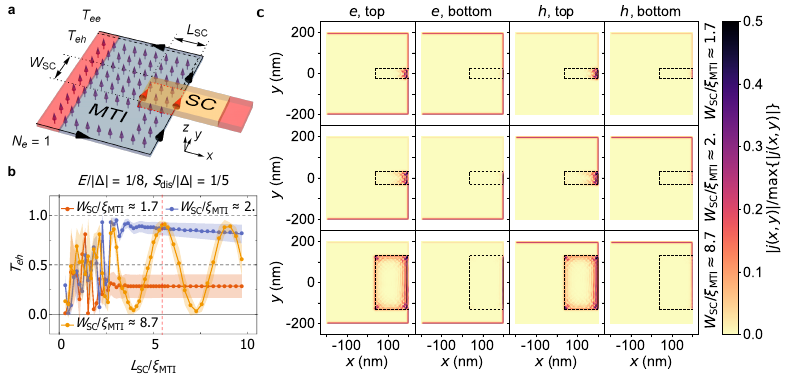}
	\caption{\linespread{1.05}\selectfont{}
		{\bf Quantum transport simulation of crossed Andreev reflection in a proximitized QAHI thin film. } 
		\textbf{a,} Schematic of the transport simulation setup with a magnetic TI (MTI) thin film in the QAHI state that is covered by a SC finger over a region with length $L_\mathrm{SC}$ and width $W_{\rm SC}$, considering the top surface of the MTI below the SC finger shifted out of the magnetic gap and proximitized into the TSC regime. The leads (red) are set to be semi-infinite. 
		\textbf{b,} The disorder-averaged electron-hole conversion probability $T_{\rm eh}$ (standard deviation indicated by shading) across the TSC region at a small bias energy $E$ is shown as a function of $L_\mathrm{SC}$
for three selected values of $W_{\rm SC}$. 
		\textbf{c,} The components of local current densities carried by electrons and holes as well as at top and bottom surfaces, plotted for three different widths of the SC finger (indicated by black dashed lines) used in {\bf b}. The finger length under consideration, indicated by vertical red dashed line in \textbf{b}, yields $T_{eh} > 0.5$ for two out of three examples, corresponding to a regime dominated by CAR in the case of a narrow finger and by the chiral Majorana edge-channel interference in the case of a wide finger.
	}
	\label{fig:simulations}
\end{figure}

An alternate scenario is that the proximitized TI surface realises a topological superconductor (TSC) region that hosts a single CM edge mode. This can happen, for instance, if the Nb slightly dopes the TI surface to make the chemical potential lie above the magnetic gap and only the top surface is proximitized \cite{Qi2010}. In this case, for a wide finger, an incoming electron hitting the TSC region splits into two CMs that take opposite paths enclosing the region covered by the finger. The two CMs recombine on the opposite side of the TSC region as either an electron or a hole, depending, in principle, on the number of residual vortices trapped in the SC region enclosed by the path \cite{Fu2009b, Akhmerov2009}. However, since the number of vortices in our several $\mu$m long Nb finger will be large and the CMs have a finite spatial extent, these processes will likely self-average, resulting in an equal mixture of electrons and holes transmitted to the opposite side of the finger due to the CM modes, such that $\Delta R_D\approx0$. On the other hand, a narrow finger, $W_{\rm SC} \lesssim \xi_s$, allows CAR processes to the opposite edge through the bulk of the proximitized TSC region and leads to $\Delta R_D<0$, as in the previous scenario of trivial SC. 
We can visualize these qualitatively different regimes of the TSC scenario in quantum transport simulations (see Fig.~\ref{fig:simulations}a) with a microscopic tight-binding model appropriate for a proximitized QAHI in the TSC regime (see Methods for details). Our simulation results presented in Fig.~\ref{fig:simulations}b and \ref{fig:simulations}c show that, when the SC finger is much wider than the induced SC coherence length, the current on the top surface is carried by CM modes traveling around the proximitized section, with the finger length and the width both affecting the interference. For example, the plot for a wide finger in Fig.~\ref{fig:simulations}b shows that the electron-hole conversion probability $T_{\rm eh}$ oscillates regularly as a function of the finger length $L_{\rm SC}$ when $L_{\rm SC} \gg \xi_s$; here, $T_{\rm eh} > 0.5$ means that predominantly holes come out on the downstream edge due to the interference of the CM modes. In a real situation with a long finger, such an oscillating $T_{\rm eh}$ self-averages to 0.5, resulting in $\Delta R_D\approx0$.  
When the finger is narrower (i.e. $W_{\rm SC} \sim \xi_s$), a qualitatively different regime is obtained: $T_{\rm eh}$ is very sensitive to $L_{\rm SC}$ for $L_{\rm SC} \lesssim \xi_s$, but it stabilizes at large $L_{\rm SC}$ to a nearly fixed value that depends sensitively on $W_{\rm SC}$. Fig.~\ref{fig:simulations}b shows the behaviour of $T_{\rm eh}$ for two different widths in the narrow regime, stabilized at large $L_{\rm SC}$ to $T_{\rm eh}$ values larger and smaller than 0.5. 
The simulated local current densities (Fig.~\ref{fig:simulations}c) suggest that there are no more well-separated CM modes in this regime and that the electron-hole conversion can be attributed to a CAR process that occurs mainly near the QAHI edge.

Therefore, our simulations suggest that, similar to the trivial SC case, the CAR process can indeed become dominant in the TSC case. 
We should nevertheless note that the stabilized value of $T_{\rm eh}$ for a narrow finger at large $L_{\rm SC}$ is strongly dependent on $W_{\rm SC}$ in our simulation and is not always larger than 0.5 (see Supplementary Note No. 10), which implies that $\Delta R_D$ would fluctuate between negative and positive as a function of $W_{\rm SC}$ in the narrow-finger regime. A similar oscillatory behaviour was also predicted by theoretical calculations for the trivial SC case~\cite{Hou2016, Beconcini2018, Galambos2022}. However, in our experiment we found $\Delta R_D$ to be always negative for narrow fingers, as was also the case with similar experiments on graphene with a trivial SC finger \cite{Lee2017, Gul2022}. This stability of negative $\Delta R_D$ in narrow fingers points to the existence of additional physics that are not captured in our simulations; in fact, the reason for the stable dominance of CAR in real experiments is an interesting subject of its own \cite{Kurilovich2022, Schiller2023, Galambos2022}. For example, the dissipative channel through vortices in the SC finger, which is not included in our simulations, could be playing a role \cite{Hatefipour2022, Schiller2023}.  In this regard, in related experiments to probe the Andreev edge states with a wide SC electrode, oscillatory $\Delta R_D$ and stably negative $\Delta R_D$ are both reported \cite{Zhao2020, Hatefipour2022}. The latter is surprising \cite{Kurilovich2023}, and possible explanations for the dominance of electron-hole conversion in the Andreev edge states have also been proposed \cite{Hatefipour2022, Tang2022, David2023, Michelsen2023}. 

One can see from the above discussions that both trivial and nontrivial scenarios are consistent with our observations. Hence, further studies will be required to determine the exact nature of the SC phase that results in CAR of the QAHI edge. Irrespective of its nature, our observation $\xi_{\rm CAR} \gg \xi_{\rm Nb}$ implies that CAR occurs through the superconductivity of the proximitized magnetic TI surface, rather than the SC finger itself. This makes sense, since Nb has negligible spin-orbit coupling and the finger on the top surface does not naturally result in processes coupling to the bottom surface, while our simulations suggest that the bottom surface needs to be involved in the CAR processes in the QAHI platform. Furthermore, the dependence on the magnetic history of the device suggests that trapped vortices and/or the magnetic disorder profile play a role,  which is natural in the above scenario for the CAR through the proximitized surface.


\begin{flushleft}
{\bf Discussion}
\end{flushleft}
\vspace{-8mm}

A negative $\Delta R_{\rm D}$ cannot be explained without the Andreev processes, and therefore our result gives unambiguous evidence that SC correlations can be induced in the chiral edge state of a QAHI. 
The stable dominance of the electron-hole conversion observed here joins the previous similar observations \cite{Lee2017, Gul2022, Hatefipour2022} and calls for better theoretical understanding.
Our experiment further shows $\xi_{\rm CAR} \gg \xi_{\rm Nb}$, which strongly suggests that in our devices superconducting correlations are induced in the 2D surface of the QAHI beneath the Nb electrode.
An obvious next step is to confirm whether the induced 2D superconductivity is topological and is associated with CM edge states. A possible experiment to address this question would be based on a device similar to ours, but has a much shorter finger electrode, such that the interference between the two CM edge states traveling along either sides of the finger can be detected without self-averaging. If the transmitted charge switches between electron and hole depending on the number of vortices in the finger, it gives strong evidence for chiral Majoranas~\cite{Fu2009b, Akhmerov2009}. Furthermore, by putting two SC fingers close enough to make a Josephson junction and by applying a voltage pulse across the junction, one can inject an edge vortex in the CM edge state. This edge vortex is a non-Abelian zero mode and experiments to confirm its non-Abelian nature have been theoretically proposed \cite{Beenakker2019a}. 
Therefore, the platform presented here offers ample opportunities to address topological superconductivity, Majorana physics, and non-Abelian zero modes. 


\begin{methods}

\subsection{Material growth and device fabrication.}

The V-doped (Bi$_x$Sb$_{1-x}$)$_2$Te$_3$ thin films were grown on InP (111)A substrates by molecular beam epitaxy (MBE) under ultra-high vacuum (UHV). High-purity V, Bi, Sb, Te were co-evaporated onto the substrate kept at a temperature of 190$^\circ$C to produce an uniform film of $\sim$8 nm thickness. The chemical potential was tuned into the magnetic gap for an optimized Bi:Sb beam-equivalent-pressure ratio of 1:4. A capping layer of 4-nm Al$_2$O$_3$ was made {\it ex-situ} with atomic layer deposition (ALD) at 80$^\circ$C using Ultratec Savannah S200 to protect the film from degradation in air. The Hall-bar devices were patterned using standard optical lithography techniques. The narrow Nb/Au superconducting contacts (45/5 nm) and the Ti/Au normal metal contacts (5/45 nm) were defined using electron-beam lithography. The Al$_2$O$_3$ capping layer was selectively removed in heated aluminum etchant (Transene, Type-D), prior to UHV sputter-deposition of the Ti/Au and Nb/Au layers. Devices A--E reported in this paper were fabricated simultaneously on the same wafer, while device F was made on a separate wafer. A clean QAHE without the need of gating was observed in all devices. Scanning electron microscopy was used to determine the width of the Nb electrodes that were covered with 5-nm-thick Au to avoid oxidization.

\subsection{Measurement setup.}
The transport measurements were performed at a base temperature of 17--25 mK in a dry dilution refrigerator (Oxford Instruments Triton 200) equipped with a 8 T superconducting magnet. All the data presented in the main text were measured using a standard DC technique with Keithley 2182A nanovoltmeters and a Keithley 2450 current source. The AC data shown in Supplementary Note 4 were measured using a standard AC lock-in technique at low frequency (3--7 Hz) using NF Corporation LI5640 and LI5645 lock-in amplifiers.

\subsection{Quantum transport simualtions.}
We have performed quantum transport simulations using the KWANT~\cite{Groth2014} package, considering a $2 \times 4$-orbital  two-dimensional tight-binding model (on a square lattice with lattice constant $a = 2$ nm), based on the following Bogoliubov-de Gennes model Hamiltonian for a proximitized magnetic TI (MTI) thin film~\cite{Yu2010, Wang2015, Chen2018}:
\begin{align}
H_\text{BdG}(k_x, k_y) &=
\begin{pmatrix}
	H_\text{MTI}(k_x, k_y) - \mu & -i\sigma_y (1 + \rho_z) \Delta/2 \\
	i \sigma_y (1 + \rho_z) \Delta^\ast/2 & \mu - H_\text{MTI}^\ast(-k_x, -k_y)
\end{pmatrix}, \\
H_\mathrm{MTI}(k_x, k_y) &= \hbar v_\mathrm{D} (k_y \sigma_x - k_x \sigma_y) \rho_z + [m_0 + m_1 (k_x^2 + k_y^2)] \rho_x + M_z \sigma_z,
\end{align}
with $\sigma_{x,y,z}$ and $\rho_{x,y,z}$ the Pauli matrices acting on the spin and pseudospin (for the top and bottom surfaces) degrees of freedom, respectively, and $\mu$ the chemical potential. We set the Dirac velocity $\hbar v_\mathrm{D} = 3$ eV  $\text{\r{A}}$, top-bottom surface hybridization $m_0 = -5$ meV and $m_1 = 15$ meV $\text{\r{A}}^2$, out-of-plane magnetization strength $M_z = 50$ meV, and proximity-induced $s$-wave pairing potential $\Delta$ (on the top surface) with $|\Delta| = 10$ meV (yielding induced SC coherence length $\xi_\mathrm{MTI} = \hbar v_\mathrm{D} / |\Delta| = 30$ nm). These model parameters yield a magnetic gap $E_{\rm gap}  = 2(M_z - |m_0|) = 90$ meV, meaning that the magnetic gap edge is located 45 meV above the Dirac point. We consider the chemical potential $\mu = 25$ meV, such that the Fermi level is nearly centred between the Dirac point and the magnetic gap edge. For obtaining a TSC in the region below the SC finger, we introduce a local shift of $\Delta\mu = 75$ meV, bringing the local Fermi level well above the magnetic gap.
Nonmagnetic (e.g., electrostatic) disorder is considered by adding a Gaussian random field to the on-site energies of the TI thin film near the position of the SC finger, characterized by disorder strength $S = 2$ meV (i.e., the standard deviation of the Gaussian) and spatial correlation length $\lambda = 10$ nm.
Note that the model parameters are not reflecting the device properties quantitatively, which would require a scattering region several orders of magnitude larger than currently considered in the simulations (in particular, due to a much larger SC finger size and induced SC coherence length). The aim is to investigate the CAR ($\xi_\mathrm{MTI} \approx W_\mathrm{SC}$) and Majorana interference ($\xi_\mathrm{MTI} \gg W_\mathrm{SC}$) regimes qualitatively.

\end{methods}

\begin{addendum}

\item[Acknowledgments:] 
We thank Y. Tokura, M. Kawasaki, and R. Yoshimi for their advice on the growth of QAHI films. We thank E. Bocquillon for fruitful discussions. This project has received funding from the European Research Council (ERC) under the European Union’s Horizon 2020 research and innovation program (Grant Agreement No. 741121) and was also funded by the Deutsche Forschungsgemeinschaft (DFG, German Research Foundation) under Germany's Excellence Strategy - Cluster of Excellence Matter and Light for Quantum Computing (ML4Q) EXC 2004/1 - 390534769, as well as under CRC 1238 - 277146847 (Subprojects A04 and B01). G.L. acknowledges the support by the KU Leuven BOF and Research Foundation–Flanders (FWO, Belgium), file No. 27531 and No. 52751.
K.M. acknowledges the support by the Bavarian Ministry of Economic Affairs, Regional Development and Energy within Bavaria’s High-Tech Agenda Project ``Bausteine für das Quantencomputing auf Basis topologischer Materialien mit experimentellen und theoretischen Ansätzen" (grant allocation no.\ 07 02/686 58/1/21 1/22 2/23), by the QuantERA grant MAGMA, and by the German Research Foundation under grant 491798118. H.F.L. acknowledges support by the Georg H. Endress Foundation.

\item[Author contributions:] A.A.T. and Y.A. conceived the project. A.U., G.L, A.B., and A.A.T. grew the thin films. G.L and L.M.C.P  analyzed the thin films. A.U. and G.L. fabricated the devices. A.U., G.L., and A.A.T. performed the experiments, and with the help from Y.A. analysed the data. K.M. performed the theoretical simulation. A.A.T. and Y.A. interpreted the data with the help of H.F.L. and K.M. A.U., Y.A., H.F.L. and K.M. wrote the manuscript with input from all authors.

\item[Competing Interests:] The authors declare no competing interests.

\item[Correspondence:] Correspondence and requests for materials should be addressed to A.A.T. (taskin@ph2.uni-koeln.de) and Y.A. (ando@ph2.uni-koeln.de).

{\bf Data availability:} The data that support the findings of this study are available at the online depository figshare with the identifier **** and Supplementary Information.

\end{addendum}

\clearpage

\begin{figure}
\centering
\includegraphics[width=\textwidth]{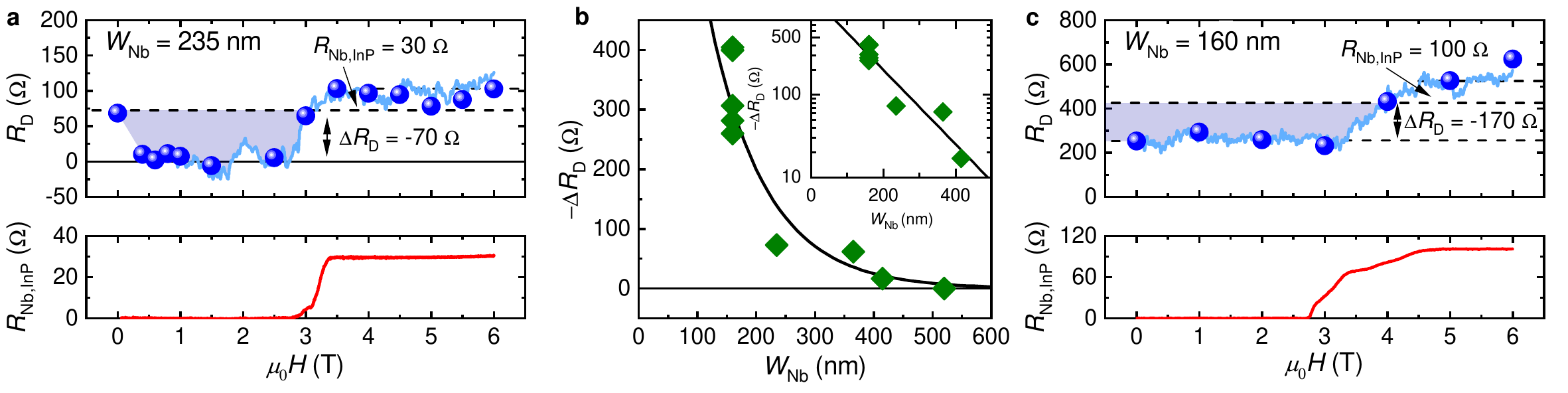}
\caption{\linespread{1.05}\selectfont{}
{\bf Dependences of the crossed Andreev reflection on the width and cleanliness. } 
\textbf{a,} Magnetic-field dependence of $R_\text{D}$ for the 235-nm-wide Nb electrode of device B shown together with $R_\text{Nb,InP}$; light blue line shows the $R_\text{D}$ continuously measured in a magnetic-field sweep at 25 mK with $I_\text{DC}=2$ nA, while blue symbols represent the slopes of the $I$-$V$ curves at $I_\text{DC}$ = 0. The $R_\text{D}$ level without $R_\text{Nb,InP}$ is marked by a dashed line. The change in the downstream resistance due to CAR, $\Delta R_\text{D}$, is estimated to be about $-70$ $\Omega$ in this sample. \textbf{b,} Exponential width dependence of $\Delta R_\text{D}$; green symbols correspond to the data for devices A--E fabricated on the same wafer (see Supplementary Note 7 for additional data). Inset shows the same data on a semi-log plot. Solid black line is a fit of the data to $\Delta R_\text{D}=R_0\exp(-W_\text{Nb}/\xi_\text{CAR})$, yielding $R_0 \approx -750$ $\Omega$ and $\xi_\text{CAR} \approx 100$ nm. \textbf{c,} Similar measurement as in panel \textbf{a} for the 160-nm-wide Nb electrode of device F, fabricated on a different wafer after a few months from the film growth. Reflecting a relatively large interface resistance between the QAHI film and Nb electrode, $R_\text{D}$ of device F is $\sim$255 $\Omega$ even when Nb is superconducting and increases to $\sim$525 $\Omega$ above $H_\text{c2}$. Nevertheless, $\Delta R_\text{D} \simeq -170$ $\Omega$ is still obtained for this 160-nm-wide Nb electrode.}
\label{fig:Width_Dep}
\end{figure}


\newpage

\end{document}


\maketitle

\section{Contact Resistance for Ti/Au contact}
 
\begin{figure}[h]
\centering
\includegraphics[width=0.8\textwidth, trim={0cm 0.3cm 0cm 0.3cm}, clip]{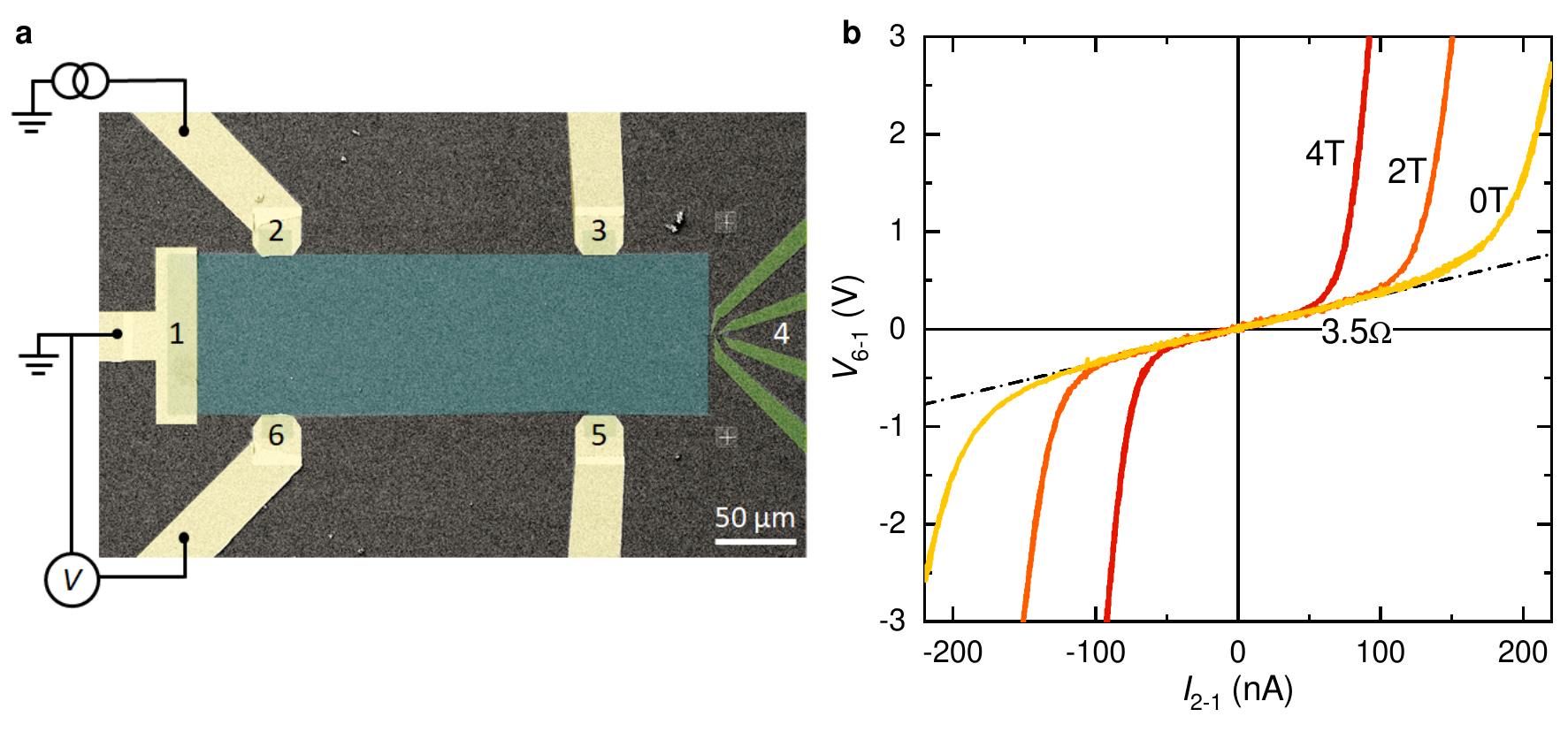}
\caption{\linespread{1.05}\selectfont{}
{\bf Three-terminal $I$-$V$ characteristics in device A for the Ti/Au contact 1. }
\textbf{a,} False-colour scanning-electron-microscope image of device A from Fig.1a, including the measurement schematics. The current flew from contact 2 to 1, and the voltage was measured between contacts 6 and 1. For an upward, out-of-plane magnetization ($M>0$), the chiral 1D edge state propagates in the counter-clockwise direction. \textbf{b,} Plots of the 3-terminal voltage $V_\text{6-1}$ as a function of the DC current $I_\text{2-1}$ for various magnetic field $H$ at 17 mK. The breakdown current of the QAHE decreases with increasing $H$. The dashed line is a linear fit to the pre-breakdown regime, yielding the slope of $\sim$3.5 $\Omega$ that corresponds to the contact resistance of the Ti/Au contact 1.}
\label{fig:Au_Contact}
\end{figure}

Figure \ref{fig:Au_Contact}a shows the schematics of the 3-terminal measurement for the contact resistance of contact 1. Figure \ref{fig:Au_Contact}b shows the voltage $V_\text{6-1}$ as a function of the DC current $I_\text{2-1}$ recorded at 17 mK in different magnetic fields. For an upward, out-of-plane magnetization, the 3-terminal (downstream) resistance $R_\text{2-1,6-1} \equiv V_\text{6-1}/I_\text{2-1}$ consists only of the sample resistance and the contact resistance. Before the current-induced breakdown of the QAHE, the sample resistance is zero. Hence, the slope of 3.5 $\Omega$ in the pre-breakdown regime is the contact resistance of the Ti/Au contact 1.

\section{Estimation of the normal-state Nb contribution}

\begin{figure}[h]
\centering
\includegraphics[width=0.8\textwidth, trim={0.3cm 0.5cm 0.3cm 0.5cm}, clip]{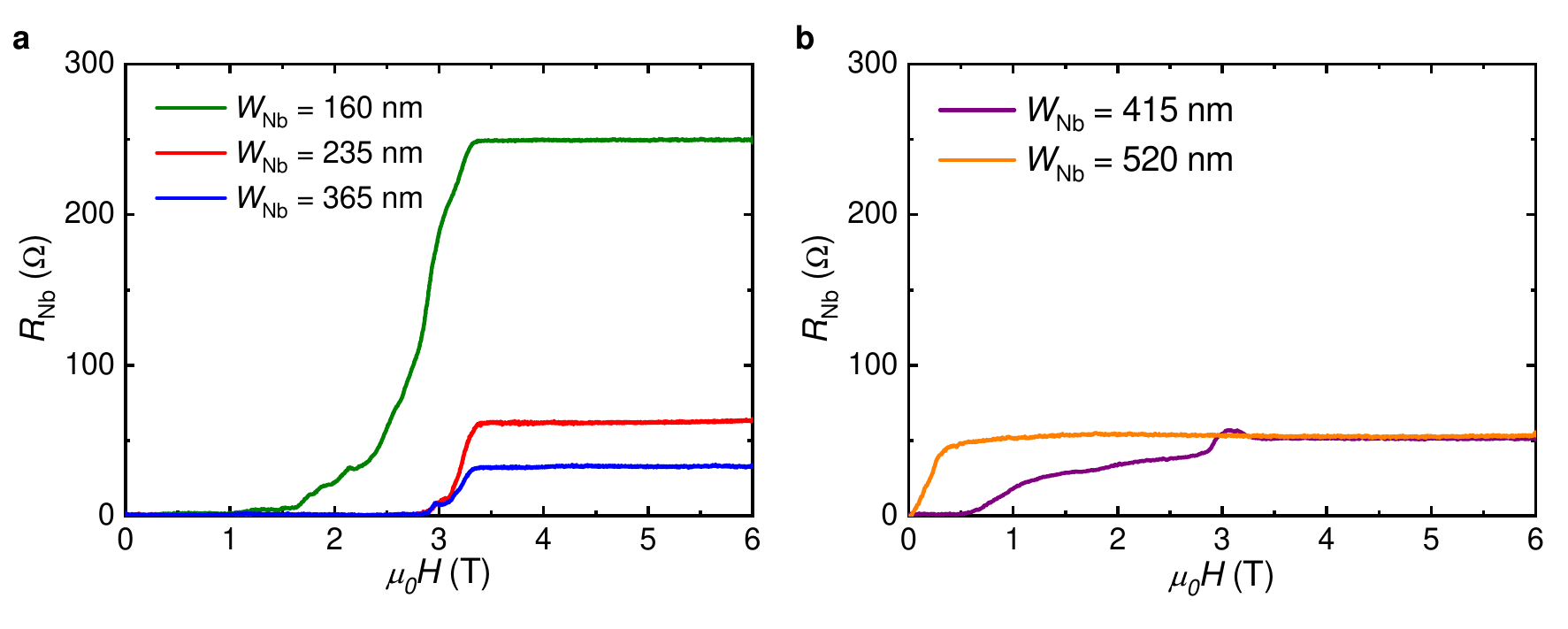}
\caption{\linespread{1.05}\selectfont{}
{\bf Magnetic-field dependence of the Nb electrode resistance. } 
\textbf{a,} The 4-terminal Nb resistance $R_\text{Nb}$ of device A ($W_\text{Nb} = 160$ nm), device B ($W_\text{Nb} = 235$ nm), and device C ($W_\text{Nb} = 365$ nm). \textbf{b,} $R_\text{Nb}$ of device D ($W_\text{Nb} = 415$ nm) and device E ($W_\text{Nb} = 520$ nm). Although the Nb electrodes on all the devices were fabricated simultaneously, the resistivity of Nb and the upper critical field $H_\text{c2}$ differ among the devices.}
\label{fig:Nb_B}
\end{figure}

In the main text, Figs. 1a-b show SEM pictures of device A. The Nb electrode contains four contacts (4a, 4b, 4c, and 4d) to allow for a 4-terminal resistance measurement, with separations $L_\text{Nb,sect.} \equiv L_\text{a-b} = L_\text{b-c} = L_\text{c-d} = 2.5$ $\mu$m. The width is $W_\text{Nb} = 160$ nm along the full length of the  Nb electrode. The overlap with the V-doped (Bi$_x$Sb$_{1-x}$)$_2$Te$_3$ thin film is $L_\text{Nb,film} = 5$ $\mu$m. The small Nb section on the InP substrate between the edge of the thin film and contact 4a has a length of $L_\text{Nb,InP} = 1.2$ $\mu$m. Devices B, C, D, and E are identical to device A except for the width of the Nb electrode: $W_\text{Nb} =$ 235, 365, 415, and 520 nm, respectively. Figure \ref{fig:Nb_B} shows the 4-terminal Nb resistance $R_\text{Nb}$ ($= V_{\rm 4b-4c}/I_{\rm 4a-4d}$) as a function of the applied magnetic field. While devices A--E are on the same wafer and the Nb electrodes were fabricated simultaneously, the resistivity of Nb and the upper critical field $H_\text{c2}$ differ among the devices. The normal-state resistance $R_\text{Nb,InP}$ of the $L_\text{Nb,InP}$-section contributes to the downstream resistance through $R_\text{D} = R_\text{QAHI} + R_\text{Nb,InP} + R_\text{SC-QAHI}$ as explained in the main text. To estimate $R_\text{Nb,InP}$ for each device, the Nb resistance is rescaled by $L_\text{Nb,InP}/L_\text{Nb,sect.}=$ 1.2 $\mu$m / 2.5 $\mu$m. This $R_\text{Nb,InP}$ was also used in the calculation of the data points for $\Delta R_\text{D}= -[R_\text{D}(H > H_\text{c2})-R_\text{D}(H < H_\text{c2})-R_\text{Nb,InP}]$ shown in Fig.3b.

\section{\textit{I}-\textit{V} characteristics at different magnetic fields in device A}
\begin{figure}[h]
\centering
\includegraphics[width=0.55\textwidth, trim={0.25cm 0.6cm 0.35cm 0.25cm}, clip]{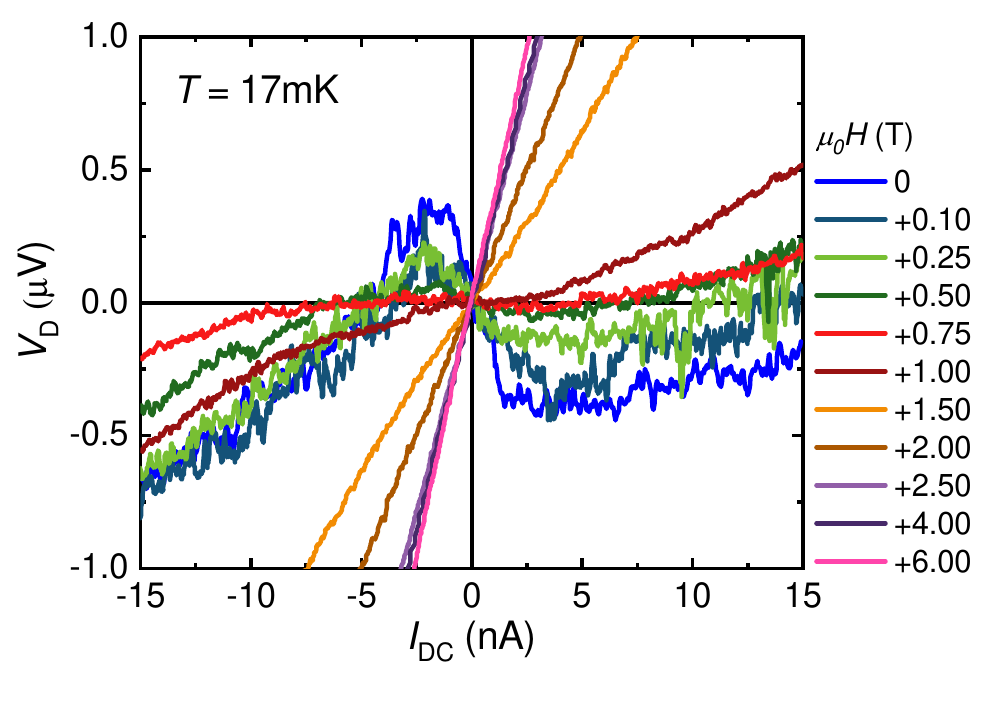}
\caption{\linespread{1.05}\selectfont{}
{\bf \textit{I}-\textit{V} characteristics at different magnetic fields in device A. } 
The negative slope in the zero-current limit, confirming the negative $R_\text{D}$, is reproducibly observed below 1 T, where the Nb electrode is still superconducting.}
\label{fig:RD_B}
\end{figure}

In the main text, Fig. 1e shows the magnetic-field dependence of $R_\text{D}$; the blue symbols represent the slopes at $I_\text{DC}$ = 0 extracted from the $I$-$V$ curves shown in Fig. \ref{fig:RD_B}. Negative $R_\text{D}$ is observed for $\mu_0 H <$ 1 T. Moreover, notice the large noise amplitude for the $V_\text{D}$-vs-$I_{\rm DC}$ curves displaying negative slopes, whereas a lower noise level is observed for curves measured above the $H_{\rm c2}$ of Nb. This indicates that the noise is intrinsic to the CAR process in this system.

\section{Comparison of \textit{R}\textsubscript{D} measured with DC and AC techniques}

\begin{figure}[h]
\centering
\includegraphics[width=0.55\textwidth, trim={0.25cm 0cm 1cm 0.25cm}, clip]{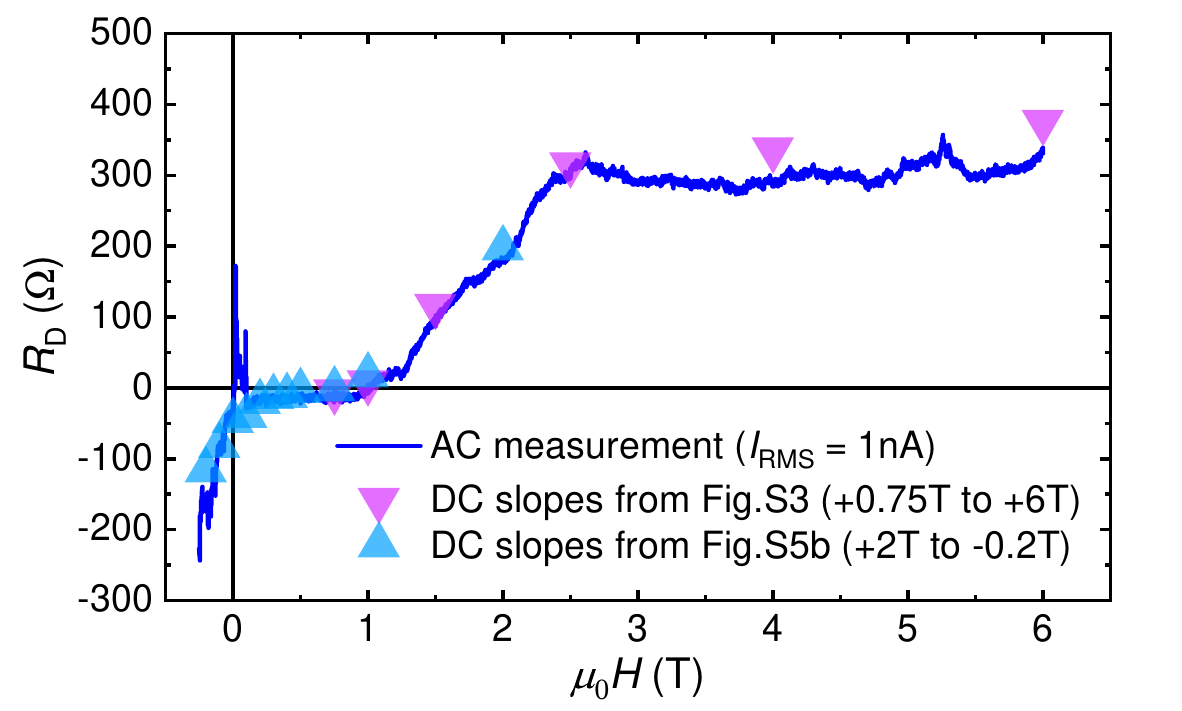}
\caption{\linespread{1.05}\selectfont{}
{\bf Validation of the \textit{R}\textsubscript{D} values in device A. } 
The solid blue line shows $R_\text{D}$ measured continuously with an AC technique with $I_\text{RMS}=\text{1 nA}$ at 25 mK as a function of the applied magnetic field; this $R_\text{D}$ result is consistent with the $R_\text{D}$ values extracted from the slopes of the $I$-$V$ curves at $I_{\rm DC}$ = 0 measured with the DC technique (magenta and cyan symbols).}
\label{fig:RD_AC}
\end{figure}

The plots of $V_\text{D}$ vs $I_{\rm DC}$ shown in Fig. \ref{fig:RD_B} were obtained with a DC technique. Up to a DC current of $|I_\text{DC}| \lesssim 3$ nA, the slope of $V_\text{D}$ (and hence $R_\text{D}$) is negative below 1 T. To verify the negative $R_\text{D}$, the sample is remeasured with an AC lock-in technique with a small AC excitation current of $I_\text{RMS} = 1$ nA (i.e. $I_\text{peak} = 1.41$ nA) for the same experimental setup as shown in Fig. 1a. Figure \ref{fig:RD_AC} shows that the result of the AC measurement agrees well with the slopes of the $I$-$V$ curves at $I_{\rm DC}$ = 0 measured with the DC technique. Hence, the negative $R_\text{D}$ in device A is reproducible between the AC and DC techniques.

\newpage 

\section{Effect of the magnetic-field-sweep history on \textit{R}\textsubscript{D} in device A}

\begin{figure}[p!]
\centering
\includegraphics[width=0.8\textwidth]{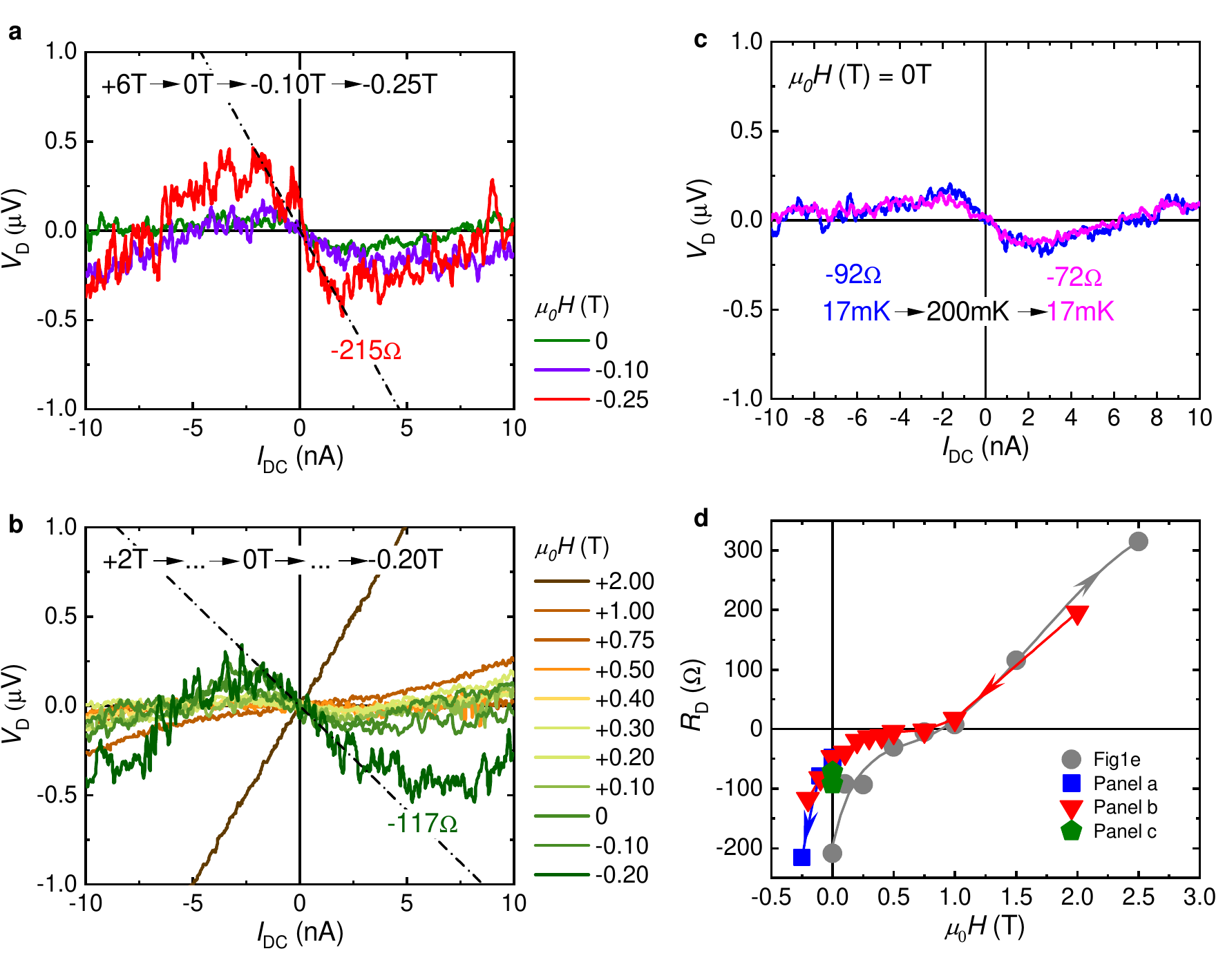}
\caption{\linespread{1.05}\selectfont{}
{\bf Effect of the magnetic-field-sweep history observed in device A.  } 
\textbf{a,} $I$-$V$ characteristics measured at 17 mK in the order of 0 , $-0.1$, and $-0.25$ T directly after taking the magnetic-field-dependence data shown in Fig. \ref{fig:RD_B}. \textbf{b,} $I$-$V$ characteristics measured at 17 mK in decreasing magnetic fields from 2 T to $-0.2$ T directly after taking the data shown in panel {\bf a} and bringing the magnetic field to 2 T. The dashed lines in panel \textbf{a} and \textbf{b} show the maximum negative slopes observed at $-0.25$ T and $-0.2$ T, respectively. \textbf{c,} $I$-$V$ characteristics measured at 17 mK in 0 T before (blue) and after (magenta) thermal cycling to 200 mK, directly after taking the data shown in panel {\bf b}. \textbf{d,} Collection of the $R_\text{D}$ values obtained in four different magnetic-field sweeps. Blue, red, and green symbols are the slopes at $I_{\rm DC}$ = 0 extracted from the data in panels \textbf{a}, \textbf{b}, and \textbf{c}, respectively. The gray symbols are the discrete data points up to 2.5 T shown in Fig. 1e.}
\label{fig:B-history}
\end{figure}

After taking the data shown in Fig. \ref{fig:RD_B} (for which $H$ was increased from 0 to 6 T), we reduced $H$ back to 0 T and took the $I$-$V$ data shown in Fig. \ref{fig:B-history}a in the order of 0 , $-0.1$, and $-0.25$ T. Then, we increased the magnetic field to 2 T and decreased it to $-0.2$ T, during which we took the series of $I$-$V$ curves at different magnetic fields shown in Fig. \ref{fig:B-history}b. Interestingly, the $R_\text{D}$ values in the zero-current limit obtained for the series shown in Fig. \ref{fig:B-history}b are essentially consistent with those obtained in the series shown in Fig. \ref{fig:B-history}a, see Fig. \ref{fig:B-history}d. This suggests that the system has metastable disorder profiles, and it remained in the same profile between the measurements of Fig. \ref{fig:B-history}a and  Fig. \ref{fig:B-history}b, while the profile changed from that in the measurements of Fig. \ref{fig:RD_B} (i.e. Fig.~1e in the main text ). The temperature dependence data shown in Fig. 2b of the main text were measured after we took the data in Fig. \ref{fig:B-history}b and set the magnetic field to zero again. 

To check for the effect of thermal cycling, we measured the 0-T $I$-$V$ curves at 17 mT before and after the sample was heated to 200 mK, and the result is shown in Fig. \ref{fig:B-history}c. It appears that the thermal cycling has little effect on $R_\text{D}$. 

To summarize the effect of the magnetic-field-sweep history in device A, Fig. \ref{fig:B-history}d shows the $R_\text{D}$ values obtained in four different magnetic-field sweeps that were performed to take the data shown in Fig. \ref{fig:RD_B}, Fig. \ref{fig:B-history}a, Fig. \ref{fig:B-history}b, and Fig. \ref{fig:B-history}c. The maximum negative slope obtained in each of Figs. \ref{fig:B-history}a--c is used as the value of $R_\text{D}$ for each magnetic cycle and we used these values to calculate $\Delta R_\text{D}$ shown in Fig. 3b of the main text.

\newpage

\section{Current- and temperature-induced breakdown of the QAHE}

\begin{figure}[h]
\centering
\includegraphics[width=0.45\textwidth, trim={0.25cm 0.25cm 0.2cm 0.25cm}, clip]{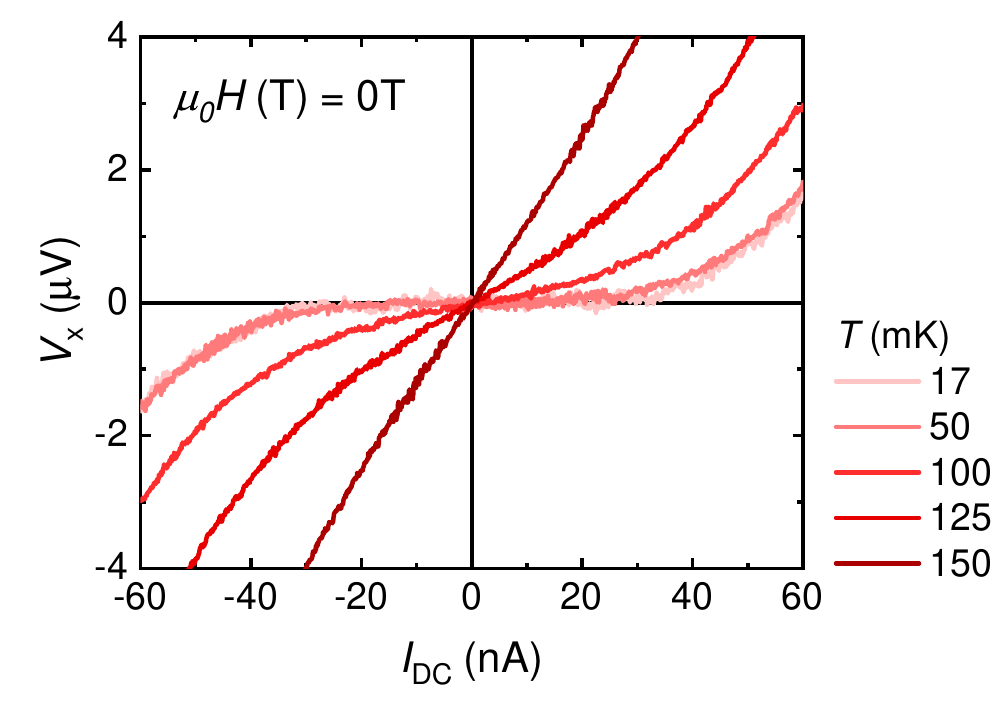}
\caption{\linespread{1.05}\selectfont{}
{\bf Breakdown of the QAHE.} 
Plots of the 4-terminal longitudinal voltage $V_x$ vs $I_\text{DC}$ in 0 T ($M>0$) measured in device A at various temperatures. The thermal activation of charge carriers with increasing temperature gives rise to a parallel dissipative conduction channel, causing the zero-resistance state of the QAHI to disappear at $\sim$100 mK, while the current-induced breakdown of the QAHE causes a finite $V_x$ above $\sim$30 nA at 17 and 50 mK.} 
\label{fig:Rxx_T}
\end{figure}

In Fig. 2b of the main text, the 4-terminal longitudinal resistance $R_\text{xx}$ is shown as a function of temperature. The data points of $R_\text{xx}$ were extracted from the $I$-$V$ curves shown in Fig. \ref{fig:Rxx_T} as the slope at $I_{\rm DC}$ = 0; here, the current was set to flow between contacts 1 and 4d, and the voltage between contacts 6 and 5 was measured. The 17-mK and 50-mK curves present a well extended zero-voltage plateau up to $\sim$30 nA, after which the current-induced breakdown of the QAHE occurs. At higher temperatures, the zero-resistance state is not realized due to the thermal activation of charge carriers into the gapped 2D surface states of the QAHI \cite{Kawamura2017, Fox2018, Fijalkowski2021, Lippertz2022}.

\newpage

\section{Downstream resistance measured with wider Nb electrodes}

\begin{figure}[h]
\centering
\includegraphics[width=1\textwidth]{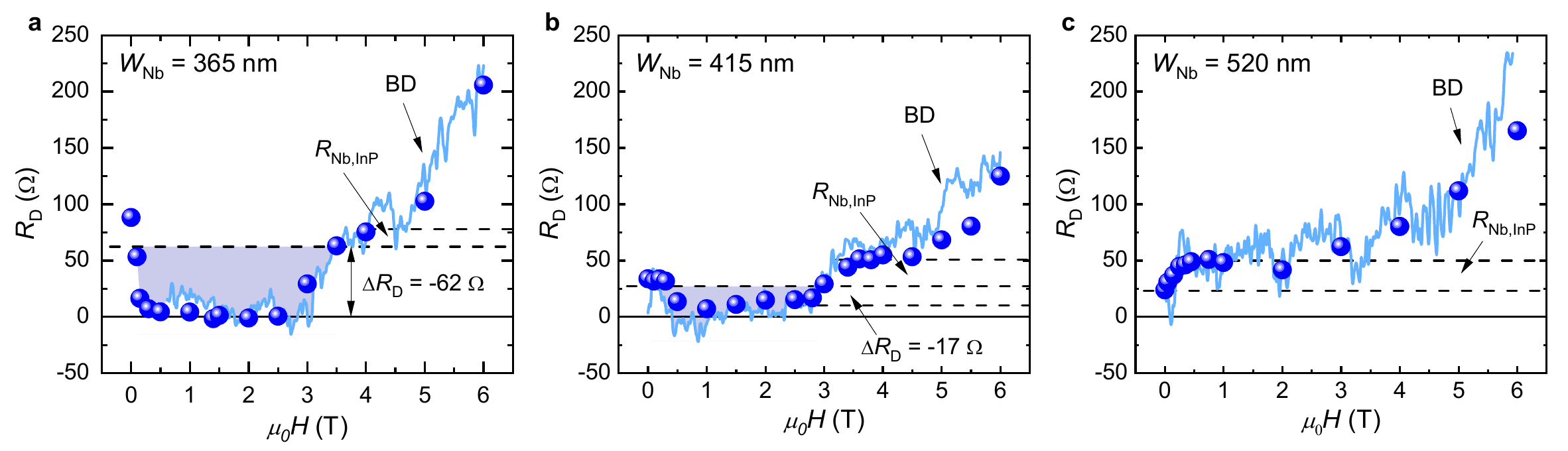}
\caption{\linespread{1.05}\selectfont{}
{\bf Downstream resistance in devices with a wider Nb electrode.} 
\textbf{a-c,} Light blue lines show the magnetic-field dependencies of $R_\text{D}$ at 25 mK measured with $I_\text{DC}=2$ nA in device C (\textbf{a}, $W_\text{Nb}=365$ nm), device D (\textbf{b}, $W_\text{Nb}=415$ nm) and device E (\textbf{c}, $W_\text{Nb}=520$ nm). Blue symbols represent the slopes in the $I$-$V$ curves at $I_\text{DC}$ = 0. Note that above $\sim$4.5 T, the breakdown of the QAHE starts to dominate $R_\text{D}$ in all these devices. The distance between two dashed lines in each panel mark the estimated normal-state Nb contribution $R_{\rm Nb, InP}$ pointed by an arrow, based on which the CAR contribution $\Delta R_{\rm D}$ is estimated.}
\label{fig:DevD+E_B}
\end{figure}

Figure \ref{fig:DevD+E_B} shows $R_\text{D}$ as a function of the applied magnetic field measured at 25 mK in devices C, D, and E having the Nb-electrode width of 365, 415, and 520 nm, respectively. Note that the increase in $R_\text{D}$ above $\sim$4.5 T observed in all devices is due to the breakdown of the QAHE, see Fig. 1d of the main text. In devices C and D, the normal-state Nb contribution was $\sim$16 $\Omega$ and $\sim$24 $\Omega$, yielding $\Delta R_\text{D}$ of about $-62$ $\Omega$ and $-17$ $\Omega$, respectively, when one compares $R_\text{D}$ before and after the superconductivity is suppressed. In device E, on the other hand, the normal-state Nb contribution of $\sim$27 $\Omega$ accounts for the full increase in $R_\text{D}$ upon the suppression of superconductivity. Hence, $\Delta R_\text{D}$ is zero for the 520-nm-wide Nb electrode of device E. The  $\Delta R_\text{D}$ values of these devices are included in Fig. 3b of the main text.

\section{Landauer-B\"uttiker formalism }

For the measurement configuration shown in Fig. 1a of the main text, the current flows from contact 1 to 4. The chiral edge state runs counter-clockwise along the sample edge for an upward, out-of-plane magnetization. The current-voltage relation in the linear-response Landauer-B\"uttiker formalism \cite{Datta1995,Lambert1998} is given by
\begin{equation}
I_i=\sum_{j=1, j\neq4}^6 a_{i j}\left(V_j-V\right),
\label{eq:LB_Ii}
\end{equation}
where $I_\text{i}$ 
is the current flowing into contact $i$, $V_\text{j}$ is the potential at contact $j$, and $V$ is the potential of the superconducting electrode (contact 4). The proportionality coefficients $a_\text{i j}$ in Eq. \ref{eq:LB_Ii} at zero temperature are given by 
\begin{equation}
a_{i j}=\frac{e^2}{h}\left(N_i^{\mathrm{e}} \delta_{i j}-T_{i j}^{\mathrm{ee}}+T_{i j}^{\mathrm{eh}}\right).
\label{eq:LB_aij}
\end{equation}
Here, $N_i^{\mathrm{e}}$ is the number of available channels for electron-like excitation in contact $i$:
\begin{equation}
N_i^{\mathrm{e}}=\sum_{j=1, j\neq4}^6\left(T_{i j}^{\mathrm{ee}}+T_{i j}^{\mathrm{eh}}\right),
\label{eq:LB_Ni}
\end{equation}
where $T_\text{ij}^\mathrm{ee}$ $(T_\text{ij}^\mathrm{eh})$ is the transmission probability of an electron from the $j$-th contact to arrive as an electron (hole) at the $i$-th contact. Since a QAHI possesses only a single chiral edge state, $N_i^{\mathrm{e}} = 1$ for all contacts. Notice that the potential difference in Eq. \ref{eq:LB_Ii} is expressed with respect to the potential of the grounded superconducting contact 4 ($V = V_4 = 0$), and the summation in Eqs. \ref{eq:LB_Ii} and \ref{eq:LB_Ni} runs only over the normal metal contacts. The non-zero transmission probabilities $T_\text{ij}$ are simply $T_\text{12}^{\mathrm{ee}}$ = $T_\text{23}^{\mathrm{ee}}$ = $T_\text{56}^{\mathrm{ee}}$ = $T_\text{61}^{\mathrm{ee}}$ = 1, $T_\text{35}^{\mathrm{ee}}$, and $T_\text{35}^{\mathrm{eh}}$.
The non-zero proportionality coefficients $a_\text{i j}$ then become  
\begin{align}
a_{11}&=a_{22}=a_{33}=a_{55}=a_{66}=\frac{e^2}{h}, \label{eq:LB_aij_2} \\
a_{12}&=a_{23}=a_{56}=a_{61}=-\frac{e^2}{h}, \label{eq:LB_aij_3} \\
a_{35}&=\frac{e^2}{h}\left(-T_{35}^\mathrm{ee}+T_{35}^{\mathrm{eh}}\right). \label{eq:LB_aij_4}
\end{align}
Using $I_\text1= -I_\text4 = I$ and $I_\text2=I_\text3=I_\text5=I_\text6=0$, Eq. \ref{eq:LB_Ii} gives a set of equations which can be solved for $I$ and $V_i$. The expression for the current and downstream resistance \cite{Galambos2022,Hatefipour2022} then become
\begin{align}
I &= \frac{e^2}{h}\left[ 1-\left(T_{35}^\mathrm{ee}-T_{35}^{\mathrm{eh}}\right)\right]V_\text{SD}, \label{eq:LB_I} \\
R_\text{D} &= \frac{V_\text{D}}{I}=\frac{h}{e^2}\frac{T_{35}^\mathrm{ee}-T_{35}^{\mathrm{eh}}}{1-\left(T_{35}^\mathrm{ee}-T_{35}^{\mathrm{eh}}\right)}, \label{eq:LB_Rd}
\end{align}
where $V_\text{SD} \equiv V_\text1-V_\text4$ and $V_\text{D} \equiv V_\text3-V_\text4$. 

$T_\text{35}^{\mathrm{ee}}$ and $T_\text{35}^{\mathrm{eh}}$ are not independent parameters, they represent transmission probabilities of an electron leaving the contact 5 and should satisfy the following relation:
\begin{align}
T_{35}^\mathrm{ee}+T_{35}^{\mathrm{eh}} + T^{\mathrm{D}}= 1, \label{eq:LB_conserv}
\end{align}
where $T^{\mathrm{D}}$ is the probability of the direct transfer of the electron into the SC contact 4. 

The $T^{\mathrm{D}}=0$ condition represents the case of a perfect superconductor, for which an electron with the energy smaller then the SC gap cannot enter SC contact directly, but only through Andreev processes with finite $T_{35}^{\mathrm{eh}}$. In the extreme case of the perfect Andreev process with $T_{35}^{\mathrm{eh}}$ = 1, one should observe a doubling of the current in the Hall bar (see Eq. \ref{eq:LB_I}) and the maximal negative downstream resistance with the magnitude of the half quantized resistance value (see Eq. \ref{eq:LB_Rd}).

The $T^{\mathrm{D}}=1$ condition represents the case when the contact 4 acts as a perfect metal, whitch can be achieved in our experiment, for example, by applying the magnetic field and fully suppressing the superconductivity in the finger. In this case, $R_{D} =0$  (see Eq. \ref{eq:LB_Rd}) as expected for an ideal metallic contact. For a real metallic contact, one would expect some finite (positive) contact resistance, which will contribute to $R_{D}$ in both metallic and SC states.

For any finite $T^{\mathrm{D}}$, the observation of a negative downstream resistance $R_{D}<0$ is a direct indication that $T_{35}^{\mathrm{ee}} < T_{35}^{\mathrm{eh}}$  (see Eq. \ref{eq:LB_Rd}), i.e. the number of holes, which arrive into the contact 3,  is larger than the number of electrons. For real samples with a relatively large contact resistance of the finger, the $T_{35}^{\mathrm{ee}} < T_{35}^{\mathrm{eh}}$ relation will be also seen as an increase in the downstream resistance $R_{D}$ when the superconductivity is supressed by the applied magnetic field.

Both $T_{35}^{\mathrm{ee}}$ and $T_{35}^{\mathrm{eh}}$ represent total probabilities for electrons and holes to get into the downstream channel after interacting with the SC finger and finally reach the contact 3. It is useful to distinguish between different contributions. In particular, crossed Andreev reflections (CAR) and direct tunneling of electrons from upstream to downstream channel (so called electron co-tunneling, CT) are expected to decay exponentially with increasing the width of the finger. We can write $T_{35}^{\mathrm{ee}} = T^{\mathrm{CT}}+T^{\mathrm{N}}$ and  $T_{35}^{\mathrm{eh}} = T^{\mathrm{CAR}}+T^{\mathrm{A}}$, where $T^{\mathrm{N}}$ and $T^{\mathrm{A}}$ represent the probabilities of all other processes at the finger to get into the downstream channel as an electron and hole, respectively. In most processes such as formation of the Andreev edge state or chiral Majorana interferometry, one would expect an equal mixture of electron and hole on a long finger, i.e. $T^\mathrm{N} = T^\mathrm{A}$.
In this case, the expression for $R_\text{D}$ reduces to
\begin{align}
R_\text{D} &=\frac{h}{e^2}\frac{T^\mathrm{CT}-T^{\mathrm{CAR}}}{1-\left(T^\mathrm{CT}-T^{\mathrm{CAR}}\right)}. \label{eq:LB_Rd2}
\end{align}
Hence, $R_\text{D}$ becomes negative for the measurement geometry shown in Fig. 1 when CAR occurs more often than CT.  Eq. \ref{eq:LB_Rd2} is also consistent with the expectation that $R_\text{D}$ goes quickly to zero with increasing width of the finger (or to a finite value in the real sample with a finite contact resistance).

\section{Wavefunction of chiral edge state}

Following similar derivations in Refs.~\citenum{Yu2010,Zhang2016}, without loss of generality and neglecting coupling to the two split-off bands far from the Fermi-level that do not have a band inversion  resulting from the magnetisation, we write the two lowest energy states as $\left|+ \uparrow_z  \right>=(\left|t \uparrow_z  \right>+\left|b \uparrow_z  \right>)/\sqrt{2}$ and $\left|- \downarrow_z \right>=(\left|t \downarrow_z  \right>-\left|b \downarrow_z  \right>)/\sqrt{2}$, where $(+)$ is a symmetric (bonding) and $(-)$ is an antisymmetric (anti-bonding) state spread over the top and bottom surface with spin $\uparrow_z$ and $\downarrow_z$ along the magnetisation axis. In the basis $(\left|+ \uparrow_z \right>,\left|- \downarrow_z \right>)$ the Hamiltonian of the lowest energy states is then given by
\begin{equation}
H=\left(\begin{array}{cc}
m_k-M & -i v (k_x+i k_y) \\
i v (k_x-ik_y) & -m_k+M
\end{array}\right)=\left(m_k-M\right) \tau_z+v\left(k_y \tau_x+k_x \tau_y\right),\label{eq:ham}
\end{equation}
where $m_k=m+B (k_x^2+k_y^2)>0$. A magnetisation $M>m$ ensures that there is a band inversion that results in the existence of the chiral edge mode.

We consider an edge state on a boundary parallel to the $x$-axis such that the state lives in the region $y>0$ and $k_x$ remains a good quantum number. For $k_x=0$ we make the Ansatz that the edge state can be expressed $\psi(y,k_x=0)=A \boldsymbol{\xi} \exp(-y/\lambda)$, which means that $1/\lambda$ has to satisfy
\begin{equation}
\left[\left(m-M-\frac{B}{\lambda^2}\right) \tau_z + \frac{i v}{\lambda} \tau_x \right]\boldsymbol{\xi} =0,
\end{equation}
which has non-trivial solutions if $\boldsymbol{\xi}=(1,\chi i)/\sqrt{2}$ with $\chi=\pm 1$ and 
\begin{equation}
\frac{1}{\lambda}= \frac{-\chi v \pm \sqrt{v^2+4B(m-M)}}{2 B}.
\end{equation}
Since physical states must decay and we consider the case where the edge state is in the region $y>0$, only $\lambda>0$ is a valid solution. Furthermore, since $m<M$, only $\chi=-1$ ensures that both spinor components always satisfy this condition.
Therefore, the edge state takes the form of Eq.~(2) of the main text:
\begin{equation}\label{eq: edge-mode}
\psi(y,k_x=0)=f(y) (\left|t \uparrow_z  \right>+i \left|t \downarrow_z  \right>) +(\left|b \uparrow_z  \right>-i\left|b \downarrow_z  \right>))/\sqrt{2}= f(y)(\left|t \uparrow  \right> +\left|b \downarrow  \right>),
\end{equation}
where $f(y)\sim \exp(-y/\lambda) $ and $\uparrow$, $\downarrow$ refers to spin in the plane of the QAHI perpendicular to the edge (here, $y$-direction).

\section{Quantum transport simulations}
\begin{figure}[h]
\centering
\includegraphics[width=0.64\textwidth]{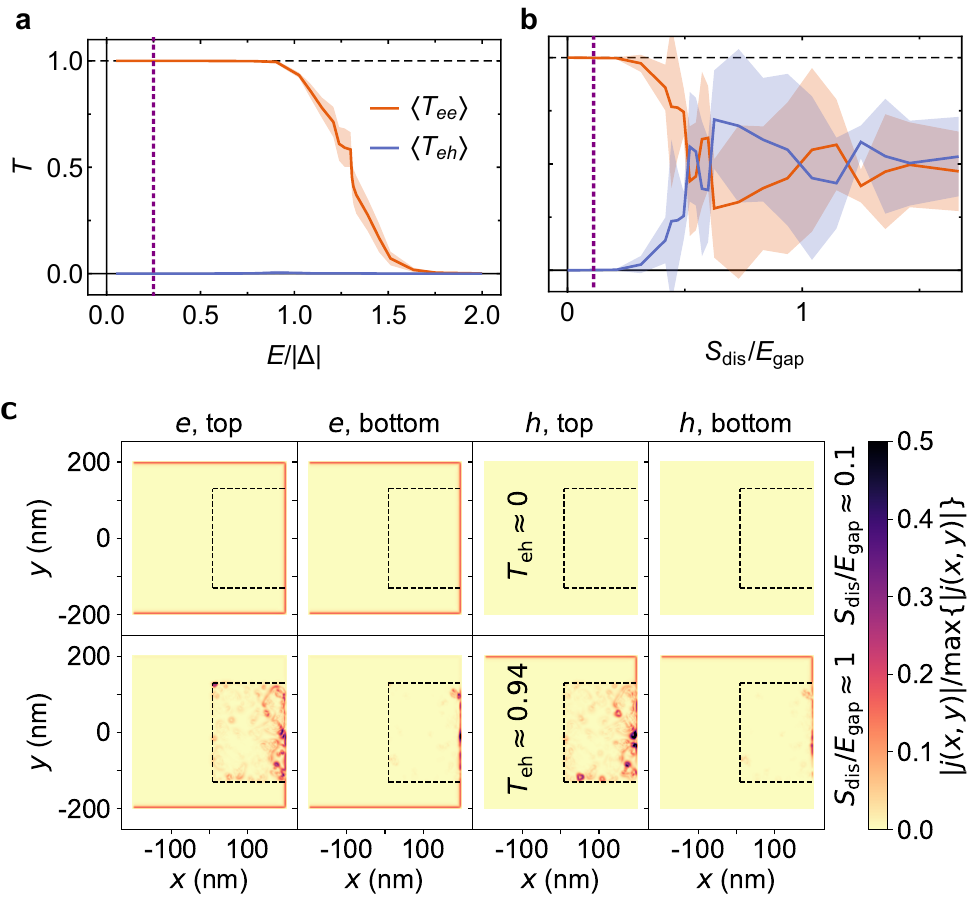}
\caption{\linespread{1.05}\selectfont{}
{\bf Quantum transport simulation of proximitized QAHI without doping from the SC finger.}
\textbf{a,b,} $\langle T_{ee} \rangle$ and $\langle T_{eh} \rangle$ obtained after averaging the results for various disorder distributions (one standard deviation is indicated by shading) shown as a function of bias energy $E$ ($|\Delta|$ = 10 meV) in \textbf{a}, and as a function of the disorder strength $S_\mathrm{dis}$ (relative to the magnetic gap $E_\mathrm{gap}$ = 90 meV) in \textbf{b}. The disorder strength (bias energy) considered in the calculations for \textbf{a} (\textbf{b}) is indicated by the dashed purple line in \textbf{b} (\textbf{a}). Here, we fixed $L_\mathrm{SC}$ = 164 nm and $W_\mathrm{SC}$ = 260 nm.
\textbf{c,} Components of the local current densities for $S_{\rm dis}/E_\mathrm{gap} = \frac{1}{9}$ (top) and $S_{\rm dis}/E_\mathrm{gap} = \frac{10}{9}$ (bottom) at the bias energy $E/|\Delta| = \frac{1}{8}$; for the latter, a disorder configuration that gave a particularly high $T_{\rm eh}$ is chosen for demonstration purpose. }
\label{fig:no-doping}
\end{figure}


Here, we present more quantum transport simulation results for the setup of Fig.~4a. 
In Fig.~\ref{fig:no-doping}, we show $T_\mathrm{ee}$ and $T_\mathrm{eh}$ as a function of the bias energy $E$ and the disorder strength $S_{\rm dis}$ for fixed $L_\mathrm{SC}$ = 164 nm and $W_\mathrm{SC}$ = 260 nm in the case of no doping, i.e., without shifting the chemical potential outside of the  magnetic gap into the TSC regime below the SC finger; the local current density distributions for two representative disorder levels are also shown. 
In this case, the top surface remains undoped and the chiral edge channel displays perfect CT at low bias and low disorder strengths. When $S_{\rm dis}$ becomes large enough to push the system locally out of the magnetic gap, electron-hole conversion starts to appear ($T_\mathrm{eh} > 0$). Note that $S_{\rm dis}$ is the standard deviation of the Gaussian random field added to the on-site energies to simulate the disorder potential. At larger $S_{\rm dis}$, the disordered region is effectively doped and $T_\mathrm{eh}$ fluctuates around 0.5 with a large standard deviation. This scenario is unlikely to apply to the experimental setup, as the QAHI state is well established in the sample. This suggests that a TSC phase due to uniform doping below the SC finger is needed in the proximitized top surface in order to mediate Andreev processes of the chiral edge channel with its peculiar spin polarization [see Eq.~\eqref{eq: edge-mode}].
Note that, with this simulation setup, we do not consider alternative (trivial) scenarios, e.g., the possibility of the QAHI edge state leaking into a subgap state in the SC finger and undergoing Andreev processes there before going back into the QAHI edge state as a hole.
Furthermore, note that we model the SC lead (shown in Fig.~4a) by a two-dimensional tight-binding model that is lattice-matched to the TI thin film model, considering parameters for a free electron gas with $s$-wave pairing ($\Delta$). It is only relevant for energies above the SC gap $|\Delta|$ when QP tunneling into the SC lead is possible, yielding $T_\mathrm{ee} + T_\mathrm{eh} < 1$ (see Fig.~\ref{fig:no-doping}a). However, in the presence of vortices, tunneling into the SC lead is possible even for $E < |\Delta|$, causing  $T_\mathrm{ee} + T_\mathrm{eh} < 1$ even at low energies.

\begin{figure}[h]
	\centering
	\includegraphics[width=\textwidth]{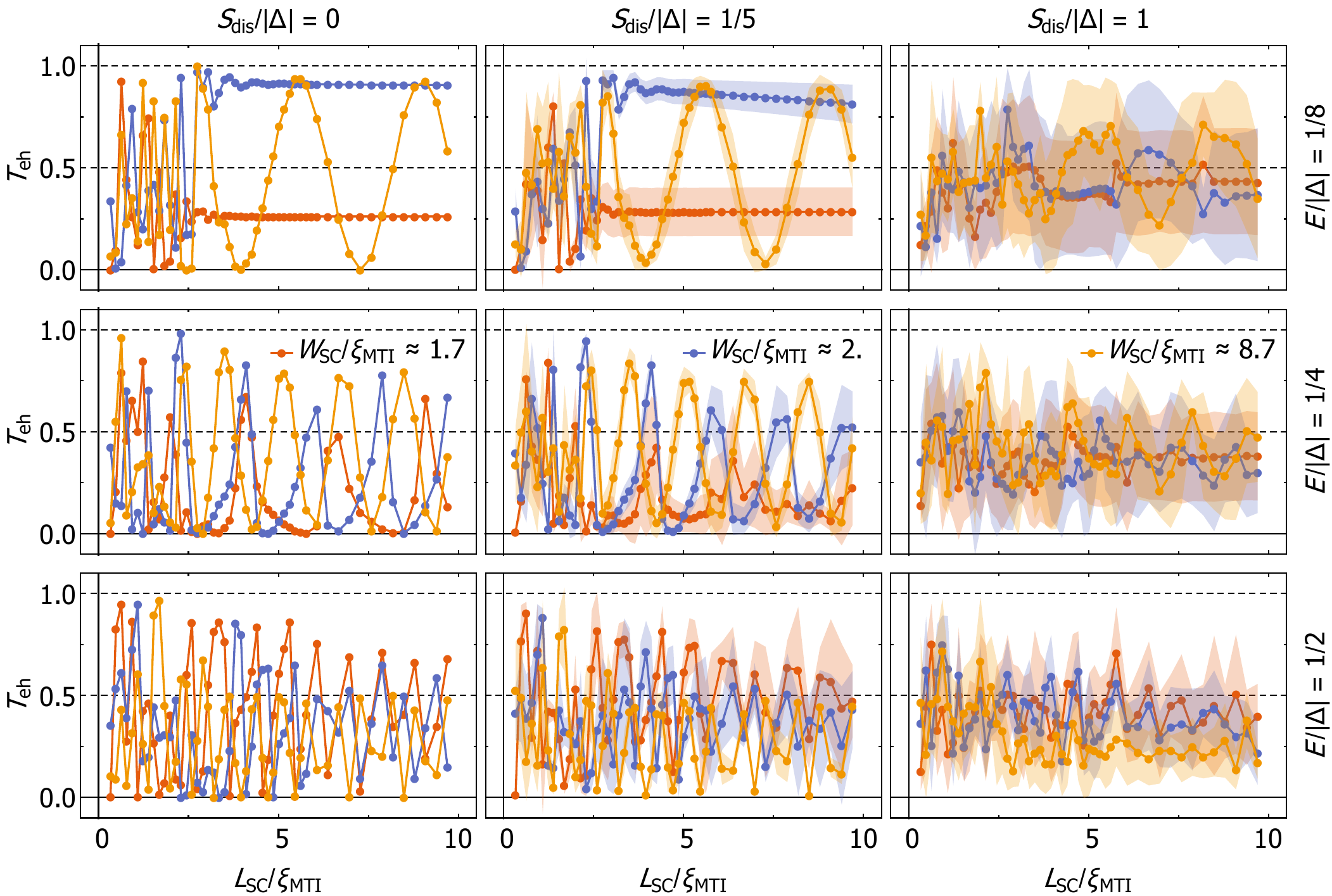}
	\caption{\linespread{1.05}\selectfont{}
		{\bf Quantum transport simulation of crossed Andreev reflection in a proximitized QAHI (extended).}
		The disorder-averaged electron-hole conversion probability $T_\mathrm{eh}$ (standard deviation indicated by shading) as a function of SC finger length as in Fig.~4b for different energies $E$ and disorder strengths $S_\mathrm{dis}$ for three selected SC finger widths (orange, blue, and dark-yellow colour correspond to $W_{\rm SC}/\xi_{\rm MTI}$ values of 1.7, 2.0, and 8.7, respectively).
	}
	\label{fig:quantum-transport}
\end{figure}

In Fig.~\ref{fig:quantum-transport}, we present $T_\mathrm{eh}$ as a function of the SC finger length $L_{\rm SC}$ as in Fig.~4b, but for different biases and disorder strengths. Increasing the disorder strength reduces the amplitude of the Majorana edge channel interference pattern around the average $T_\mathrm{eh} \leq 0.5$, whereas increasing the bias energy pushes down the amplitude of the interference pattern towards $T_\mathrm{eh} = 0$. This indicates that, for obtaining a clean Majorana edge-channel interference pattern in wide fingers and for identifying a qualitatively different CAR/CT-dominated regime in narrow fingers, disorder and bias should be sufficiently small compared to the magnetic gap and the proximity-induced pairing potential. 

Although our quantum transport simulations for narrow SC fingers indeed support the possibility of CAR to take place in the QAHI edge, they do not yield a regime with $T_\mathrm{eh} > 0.5$ that is robust against small variations in the setup (e.g., finger width or bias energy). This is different from experiment. As we mentioned in the main text, there should be additional physics which causes the stable dominance of CAR in real situation. One such possibility is the dissipation into subgap states in the SC finger (not included in our simulation setup, where $T_\mathrm{ee} + T_\mathrm{eh} = 1$ is assumed for all subgap energies $E < |\Delta|$), as suggested in Ref.~\citenum{Hatefipour2022}. When tunneling of electrons into the SC (which leads to dissipation) is allowed at low energies in addition to the tunneling into the downstream edge, the CT process would compete more with such a tunneling process than CAR, yielding $\langle T_\mathrm{eh} \rangle > \langle T_\mathrm{ee} \rangle$ even when CAR and CT are equally likely in the case without dissipation. We can further speculate that such an (imbalanced) dissipative process only starts to appear when the SC finger is narrow enough, because for wide fingers the chiral Majorana edge channels are well formed and they may short-circuit the tunneling processes.

\section*{References:}